\documentclass[sigconf]{acmart}
\AtBeginDocument{%
  }

\copyrightyear{2026}
\acmYear{2026}
\setcopyright{cc}
\setcctype{by}
\acmConference[SenSys '26]{ACM/IEEE International Conference on Embedded Artificial Intelligence and Sensing Systems}{May 11--14, 2026}{Saint Malo, France}
\acmBooktitle{ACM/IEEE International Conference on Embedded Artificial Intelligence and Sensing Systems (SenSys '26), May 11--14, 2026, Saint Malo, France}
\acmDOI{10.1145/3774906.3802764}
\acmISBN{979-8-4007-2309-4/2026/05}
\usepackage{multirow}

\usepackage{url}
\usepackage{subcaption} 

\usepackage[ruled,vlined,linesnumbered]{algorithm2e}
\newcommand{\app}{\textsc{EMPalm}}
\definecolor{darkgreen}{RGB}{34,139,34}
\newcommand{\chTransmission}{\textit{Interleaved Dual-Modal Emissions}}
\newcommand{\chLocalization}{\textit{Noisy Wide-band Spectrum}}
\newcommand{\chBit}{\textit{Bit-Level Grayscale Collisions}}
\newcommand{\chTexture}{\textit{Degraded Image Texture}}

\newcommand{\etc}[0]{\textit{etc.}\ } 
\begin{document}

\title{EMPalm: Exfiltrating Palm Biometric Data via Electromagnetic Side-Channel}

% \affiliation{%

\author{Haowen Xu$^1$, Tianya Zhao$^2$, Xuyu Wang$^2$, Lei Ma$^1$, Jun Dai$^{1\dagger}$, \\ Alexander Wyglinski$^1$, Xiaoyan Sun$^{1\dagger}$}

\affiliation{%
  \institution{$^1$Worcester Polytechnic Institute, USA; $^2$Florida International University, USA} \city{} \country{}
}

\email{Emails: {hxu4, lma5, jdai, alexw, xsun7}@wpi.edu, {tzhao010, xuywang}@fiu.edu}

\thanks{$\dagger$ Corresponding Authors}
% \author{Haowen Xu} \author{Tianya Zhao} \author{Xuyu Wang} \author{Lei Ma} \author{Jun Dai} \author{Alexander Wyglinski} \author{Xiaoyan Sun} \authornotemark[2]

% \authornote[2]{Xiaoyan Sun is the corresponding author.}

% \affiliation{%
%   \institution{$^1$Worcester Polytechnic Institute, USA; $^2$Florida International University, USA}
% }

% \email{Emails: {hxu4, lma5, jdai, alexw, xsun7}@wpi.edu, {tzhao010, xuywang}@fiu.edu}

% \affiliation{%

% \affiliation{%
\renewcommand{\authors}{Haowen Xu, Tianya Zhao, Xuyu Wang, Lei Ma, Jun Dai, Alexander Wyglinski, Xiaoyan Sun}
\renewcommand{\shortauthors}{Xu et al.}

\begin{abstract}
Palm recognition has emerged as a dominant biometric authentication technology in critical infrastructure. These systems utilize palm-related biometric features, including palmprint and palmvein data, either individually in a single-modal setting or jointly in a dual-modal.
Despite the different forms, they all employ similar hardware architectures that inadvertently emit electromagnetic (EM) signals during operation.
%Despite this diversity, they share similar hardware architectures that inadvertently emit electromagnetic (EM) signals during operation.
Our research reveals that these EM emissions leak palm biometric information, motivating us to develop \app{}—an attack framework that covertly recovers both palmprint and palmvein images from eavesdropped EM signals. Specifically, we first separate the interleaved transmissions of the visible (palmprint) and NIR (palmvein) modalities, identify the informative frequency bands of each modality, and then combine these bands to reconstruct the corresponding images. To overcome the strong noise and distortions inherent in side-channel acquisition, we further employ a diffusion model to restore fine-grained biometric features. Evaluations on seven prototype and three commercial palm acquisition devices show that \app{} can recover biometric information from real human palms with high visual fidelity, achieving Structural Similarity Index Measure (SSIM) scores up to 0.79, Peak Signal-to-Noise Ratio (PSNR) up to 29.88 dB, and Fréchet Inception Distance (FID) scores as low as 6.82 across all tested devices. Compared with the best state-of-the-art method, which can only reconstruct palm-vein images, \app{} improves overall reconstruction fidelity by 33\% and uniquely supports high-quality recovery for both palmprint and palm-vein modalities. To assess the practical implications of the attack, we further evaluate the recovered palm images against four state-of-the-art palm recognition models through real-time experiments, achieving a model-wise average spoofing success rate of 65.30\%.
\end{abstract}





\maketitle

\section{Introduction}

Palm recognition technologies, encompassing unimodal approaches based on palmprint or palmvein and multimodal methods that fuse the two, have rapidly emerged as highly secure and reliable biometric authentication techniques~\cite{8283612,gao2025deep}. In particular, multimodal fusion of palm textures with vascular structures yields high entropy, strong forgery resistance, and lasting physiological stability~\cite{20241arge}. 
Consequently, palm-based authentication has been widely adopted across government and commercial sectors, including the FBI, the Department of Homeland Security, Amazon, and Tencent~\cite{NGI,zhong2019centralized}.

% In 2013, the FBI launched the National Palm Print System and incorporated palmprint data into its Next Generation Identification platform. Similarly, the Department of Homeland Security integrates palm biometrics into its HART database. On the commercial side, companies such as Amazon and Tencent have deployed palm recognition for payments~\cite{zhong2019centralized}, underscoring its growing real-world adoption.

Traditional image-based palm recognition systems rely on either palmprint or palmvein imaging, using visible light for palmprint textures and near-infrared (NIR) sensing for subcutaneous veins~\cite{bowyer2016handbook}. Since single-modal approaches are often affected by environmental or physiological factors, modern systems overcome these limitations by adopting dual-mode architectures that capture both features simultaneously~\cite{SUNNY,2011joint,20241arge,deptrum}, thus improving accuracy and robustness. However, in both single- and dual-mode designs, sensor circuits carry time-varying currents that, by Maxwell’s equations~\cite{maxwell1890scientific}, inevitably emit electromagnetic (EM) radiation. In addition, high-speed transmission of biometric images over buses or flat cables can turn wiring into unintended antennas, exposing sensitive information through EM emissions. 

\begin{figure}[t]
    \centering
    % (a) Overall attack scenario of APP first
    \subcaptionbox{Overall attack scenario of \app{}.\label{attack_scenario}}{%
        \includegraphics[width=0.95\linewidth]{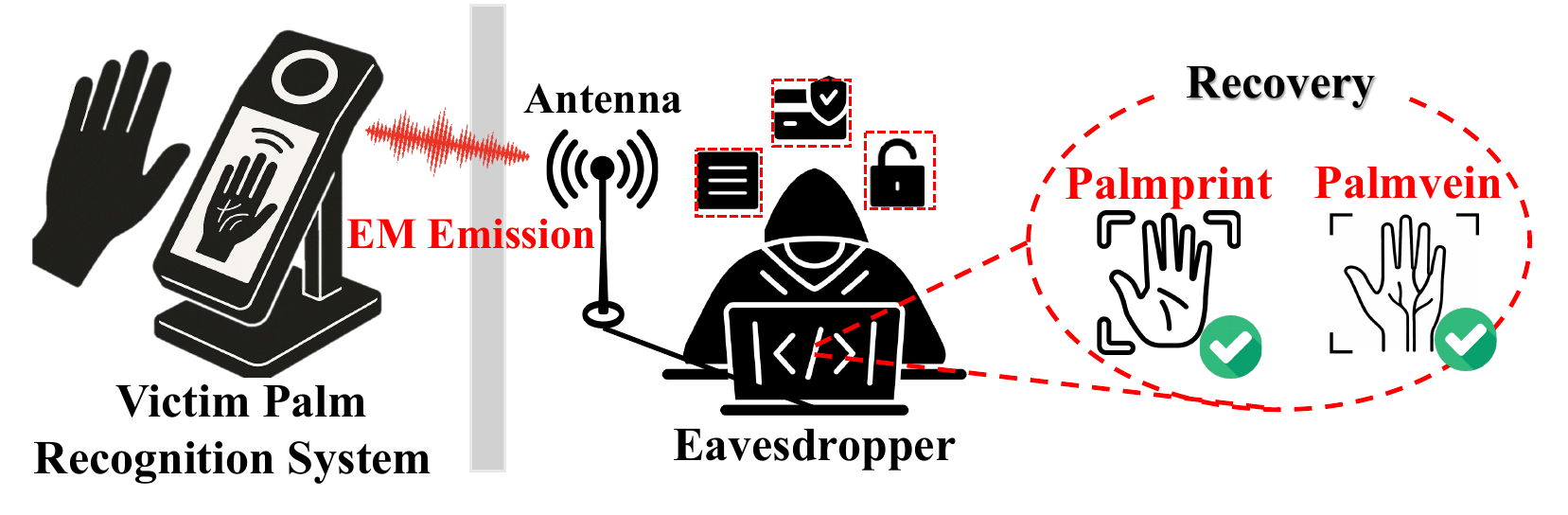}
    }
    \vspace{6pt} % 调整上下间距
    % (b) Attacking a commercial device second
    \subcaptionbox{Attacking a commercial palm recognition device using a concealed eavesdropper.\label{fig:real}}{%
        \includegraphics[width=0.8\linewidth]{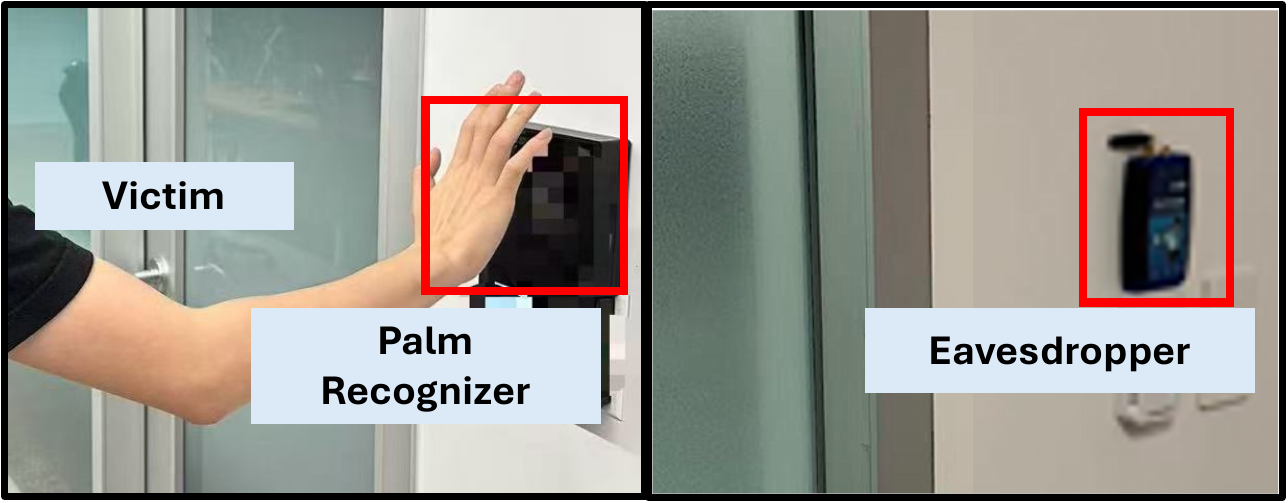}
    }
    \vspace{-1em}
    \caption{Attack scenarios of \app{}. (a) illustrates the overall attack setup and workflow, while (b) demonstrates a real-world case where an attacker covertly captures EM emissions from a commercial palm recognition device.}
    \label{fig:attack_scenarios}
    \Description{This figure shows }
\end{figure}

% Although prior research on EM leakage,such as iris recognition and embedded cameras, provide useful insights, EM leakage in palm recognition—particularly in dual-mode designs—has received little attention. Building on this gap, we demonstrate that biometric image data in palm recognition systems, including both single- and dual-mode architectures, can be eavesdropped via EM side channels. 
Although prior studies on EM leakage in biometric contexts such as fingerprint sensors~\cite{ni2023finger} and iris recognition~\cite{emiris} have provided valuable insights, EM leakage in palm recognition systems, particularly in dual-modal designs, has received limited attention. This gap is increasingly important as palm recognition is being deployed more widely for secure access control and payment authentication due to its rich biometric features and built-in liveness properties, with adoption extending to national intelligence agencies~\cite{dpa2018bndmove} and major financial institutions~\cite{amazonOne2023}. To demonstrate this, we show that biometric image data in palm recognition systems can be eavesdropped via EM side channels. As illustrated in Figure~\ref{attack_scenario}, an eavesdropper can covertly capture EM emissions from a palm scanner and reconstruct palm images as the victim performs identification, while the victim remains unaware. To our best knowledge, \app{} is the first to investigate EM leakage in palm recognition systems, and introduces the first technique capable of separating and reconstructing dual-modal biometric streams transmitted in image-based palm recognition systems.

\noindent\textbf{Challenges. }An effective eavesdropping of palm recognition systems faces four key challenges.  

\begin{itemize}
\item \textit{Interleaved Dual-Modal Emissions.} Palmprint and palmvein data can be transmitted in an alternating fashion, producing interleaved emissions that complicate modality separation.  

\item \textit{Noisy Wide-band Spectrum}. EM emissions span wide and device-dependent frequencies, making it nontrivial to identify biometric-relevant bands.

\item \textit{Bit-Level Grayscale Collisions}. Bit-packed formats cause multiple grayscale values to map to identical EM patterns, collapsing subtle intensity differences and fine details.  

\item \textit{Degraded Image Texture}. Reconstructed images exhibit degraded textures due to EM interference, environmental noise, and information loss during reconstruction.
\end{itemize}

\noindent\textbf{Our Approach.} In this paper, we present \app{}\footnote{EMPalm Project is available at \url{https://github.com/submission695-ai/Submission}}, the first EM side-channel eavesdropping attack that recovers both high-quality palmprint and palmvein from palm-recognition systems. Using unintentional EM emissions collected from palm recognition systems, \app{} recovers preliminary biometric data through a multi-stage reconstruction pipeline. To address the challenge of \chTransmission{}, we reverse-engineer transmission protocols and implement frame boundary detection, modality classification, and signal disentanglement for synchronized palmprint–palmvein reconstruction. To cope with the \chLocalization{}, we design a rapid localization framework that integrates spectrum analysis, temporal profiling, and device characterization to identify informative frequency bands. To resolve \chBit{}, we introduce a multi-band image combination strategy that leverages higher-order harmonics to restore collapsed intensity variations and preserve fine details. Finally, to mitigate \chTexture{}, we formulate the task as image restoration and employ a structure-guided diffusion model to recover high-fidelity palmprint creases and palmvein patterns.

Evaluated on seven prototype and three commercial palm recognition devices with \textbf{\textit{real human hands}}, \app{} achieves high-fidelity reconstruction with an average Structural Similarity Index Measure (SSIM) of 0.68, Peak Signal-to-Noise Ratio (PSNR) of 24.1 dB, and Fréchet Inception Distance (FID) of 8.7. Compared with state-of-the-art (SOTA) frameworks~\cite{emiris,long2024eye}, \app{} consistently delivers higher reconstruction quality and visual realism, achieving a 33\% improvement in SSIM and enhanced spoofing effectiveness against palm recognition models under identical evaluation conditions. When evaluated against four state-of-the-art practical palm-recognition models, the reconstructed images reach an average spoofing success rate of 65.3\%, confirming the practical effectiveness of the recovered biometrics.

% The diffusion model, trained on 24,702 palmprint images from 700 subjects and 11,000 palmvein images from 550 subjects, enhances preliminary reconstructions into coherent images with high perceptual fidelity, achieving an average SSIM of 0.71 and FID of 9.37. To further assess attack effectiveness, we evaluate \app{} against three palmprint recognition models and one palmvein model. When the restored images are input into these targets, they enable high-fidelity spoofing, reaching a model-wise average success rate of 65.30\% across 6,000 samples from 100 distinct users.  

\noindent\textbf{Ethical consideration. }This study was approved by the Institutional Review Board (IRB) of the participating institution, ensuring compliance with ethical and privacy standards in volunteer recruitment and data collection. We anonymized all personal information and withheld specific device models to maintain confidentiality and give vendors time to address the identified vulnerabilities.

\noindent\textbf{Contributions.} In summary, our contributions are as follows:

\begin{itemize}
    \item \textit{EM Side-channel Attack Surface Exploitation.} We first reveal EM leakage in palm biometric recognition, enabling effective spoofing of recognition models and exposing the feasibility of physical attacks.
    
    \item \textit{End-to-End Attack Framework. }We propose an end-to-end framework that includes frequency localization, single-band reconstruction, multi-band combination, and diffusion-based restoration, demonstrating robust eavesdropping capability against both single and dual modal palm recognition systems.
    
    % without requiring large paired datasets of palmprint and palmvein.
    
    \item \textit{Comprehensive Experimental Evaluation.} The effectiveness of \app{} is validated through real-world experiments on human subjects across seven prototype and three commercial palm-acquisition devices, evaluated against four state-of-the-art recognition models. \textit{Single- and dual-modal restoration} demonstrates that intercepted EM emissions can reliably recover both palmprint and palmvein modalities. \textit{Spoofing efficacy (1:100 identification)} shows that reconstructed and diffusion-enhanced images can successfully deceive advanced recognition systems. \textit{Robustness analyses} further confirm attack viability across diverse distances, orientations, intervening materials, and hardware platforms.
\end{itemize}
    
    %(including varying SBC platforms, LNA amplification levels, environmental conditions, stand-off distances up to 4 m, probe orientations, building materials, EM shielding and deferred attack)

% \begin{figure}
%     \centering 
%     \includegraphics[width=\columnwidth]{FIG/attack scenario.pdf}
%     \caption{Attack scenario of \app{}.} 
%     \label{attack_scenario}
% \end{figure}

% 在正文中
%-------------------------------------------------------------------------------
\section{Preliminaries}

\subsection{Image-based Palm Recognition}

\begin{figure}[t]
    \centering % 居中整个 figure
    
    % --- 子图 (a) ---
    \begin{subfigure}[b]{\columnwidth} % 让子图宽度等于列宽
        \centering
        \includegraphics[width=\linewidth]{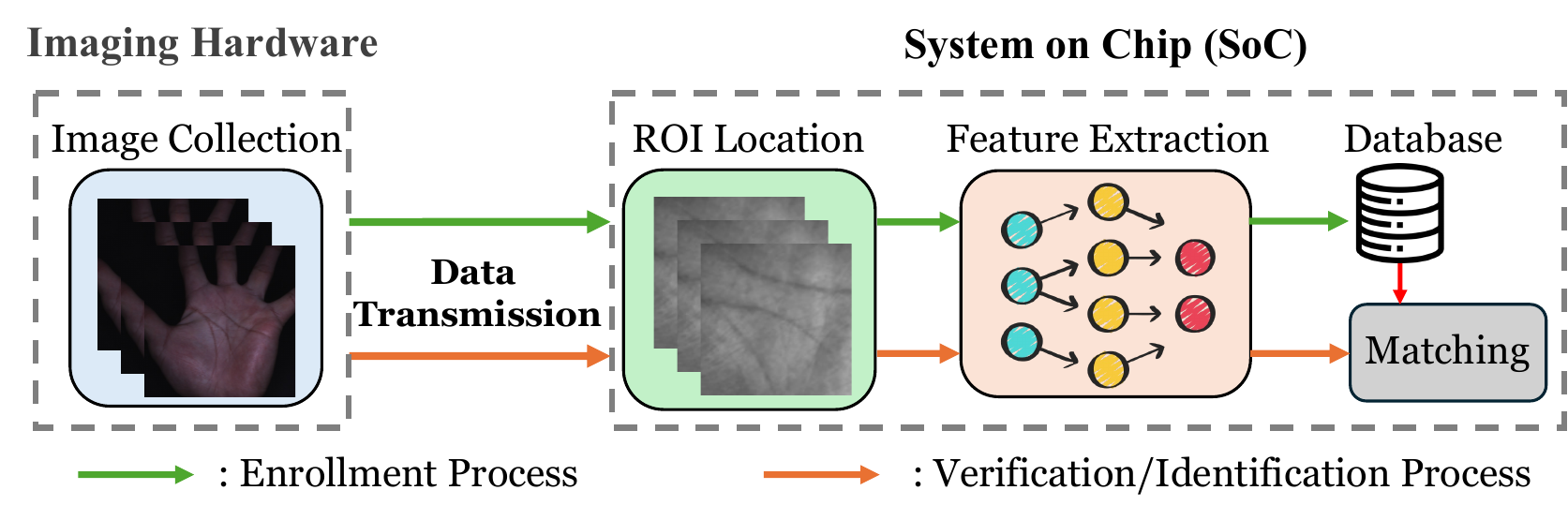}
        \caption{General workflow of palm recognition.}
        \label{workflow} % 你原来的 label
    \end{subfigure}
    % 在两个子图之间添加一点垂直间距

    % --- 子图 (b) ---
    \begin{subfigure}[b]{\columnwidth} % 让子图宽度等于列宽
        \centering
        \includegraphics[width=\linewidth]{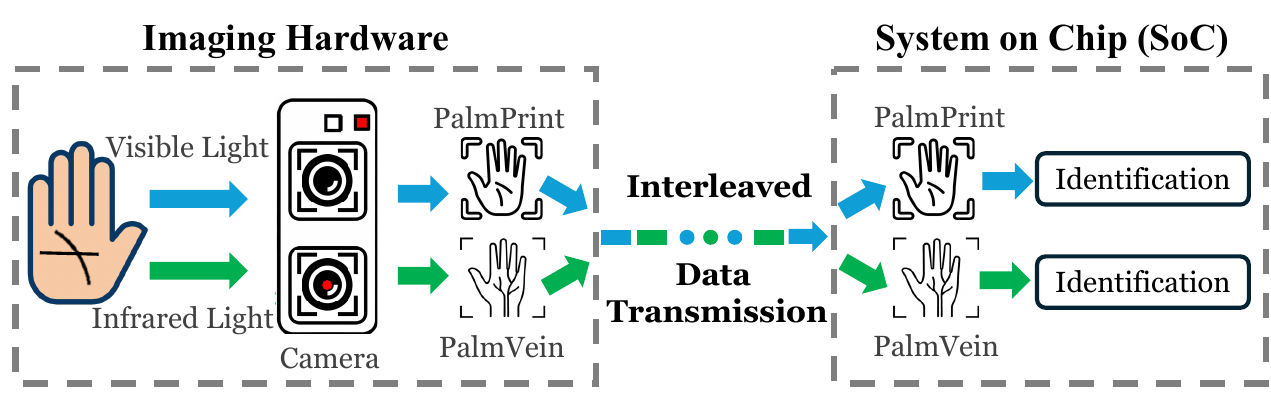}
        \caption{Workflow of a dual-modal palm recognition system. }
        \label{dual} % 你原来的 label
    \end{subfigure}
    \vspace{-2em}
    \caption{Workflow of palm recognition systems. } 
    \label{fig:system_overview} % 为整个 figure 设置一个新的 label
    \vspace{-2em}
    \Description{This figure shows ...}
\end{figure}

Figure~\ref{workflow} depicts the standard palm recognition pipeline, including image acquisition, Region of Interest (ROI) localization, feature extraction, and matching. Palm images are first captured by the imaging hardware, after which ROI localization is performed on the System on Chip (SoC) to support reliable feature analysis~\cite{ROI2005histograms}. Extracted features are then used for enrollment or compared against stored templates for verification (1:1) and identification (1:N). This pipeline applies to both palmprint and palm vein recognition systems and remains the dominant paradigm in camera-based implementations. While recent work such as mmPalm~\cite{mmpalm} explores mmWave-based palm recognition, our work investigates EM vulnerabilities in conventional imaging-based systems.

\noindent\textbf{PalmPrint Recognition.} Palmprint recognition~\cite{kong2009survey} utilizes the surface-level features of the human palm, such as principal lines and wrinkles, to perform identity verification.  The field has evolved from early statistical methods to modern deep learning approaches~\cite{zhang2012comparative}, significantly improving recognition accuracy and robustness. 
% A critical component of a palmprint recognition system is the extraction of the ROI, which isolates the central palm area containing the most discriminative features. 

\noindent\textbf{PalmVein Recognition.} Palmvein recognition~\cite{kang2014contactless} captures the internal vascular structure of the palm using NIR imaging technology. By relying on subcutaneous vascular patterns rather than the superficial skin textures used in palmprint recognition, palmvein recognition achieves greater stability and robustness, being less affected by external conditions such as skin dryness, scars, \etc.

\noindent\textbf{Dual-Modal Palm Recognition System.} 
Modern palm recognition systems increasingly adopt dual-modal architectures~\cite{SUNNY,2011joint,20241arge,deptrum} that jointly capture palmprint and palm vein information to improve accuracy and security. As shown in Figure~\ref{dual}, these systems follow the standard biometric pipeline of image acquisition, ROI localization, feature extraction, and matching. Unlike single-modal designs, visible and infrared images are acquired and transmitted as interleaved streams to the SoC for decoding and ROI extraction. The two modalities are processed independently for identity verification~\cite{zhang2018palmprint}, and their matching results are fused at the decision level to enhance robustness against spoofing and environmental variations.
% To ensure the generalizability and practicality of our proposed reconstruction approach, we conduct experiments on both unimodal systems, which capture only palmprint or only palm vein, and dual modal systems, which interleave the two modalities through a shared transmission channel\lei{right place?}.

\subsection{Image Transmission Principles}
In embedded image acquisition, sensors generate RAW images containing unprocessed pixel data from a single color component defined by the front-end filter array. These RAW images are then transferred to the image signal processor (ISP) through high-speed serial links~\cite{gunturk2005demosaicking}, most notably the MIPI Camera Serial Interface 2 (MIPI CSI-2)~\cite{mipi2023csi2}, where the debayering process is applied to interpolate missing color values of each pixel based on spatial correlations with surrounding pixels.

\begin{figure}[t]
    \centering
    % --- 子图A ---
    \begin{subfigure}{0.61\columnwidth}
        \centering
        \includegraphics[width=\linewidth]{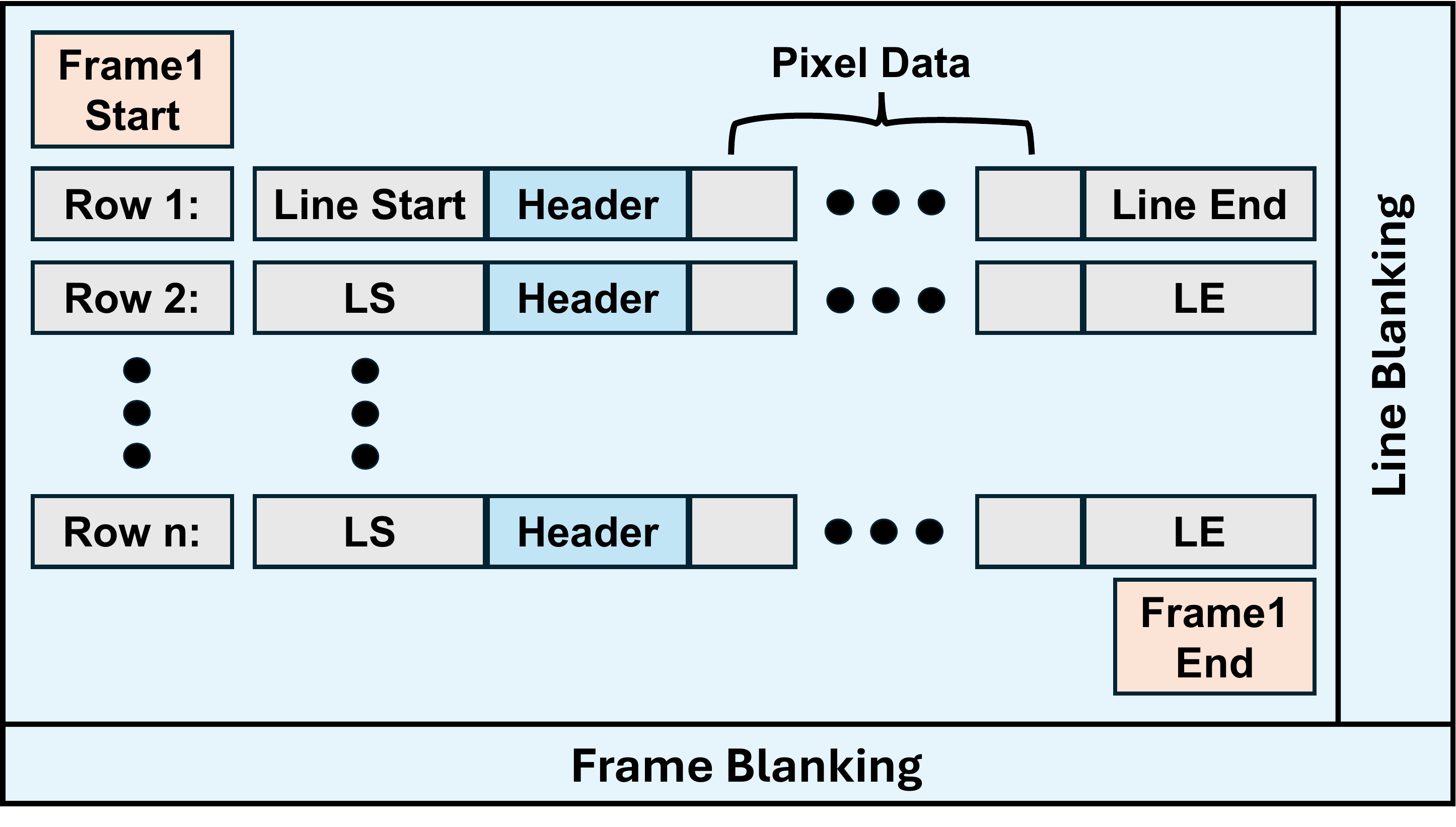}
        \caption{}
        \label{fig:sub_a}
    \end{subfigure}
    \hfill
    % --- 子图B ---
 \begin{subfigure}{0.36\columnwidth}
        \centering
        \includegraphics[width=\linewidth]{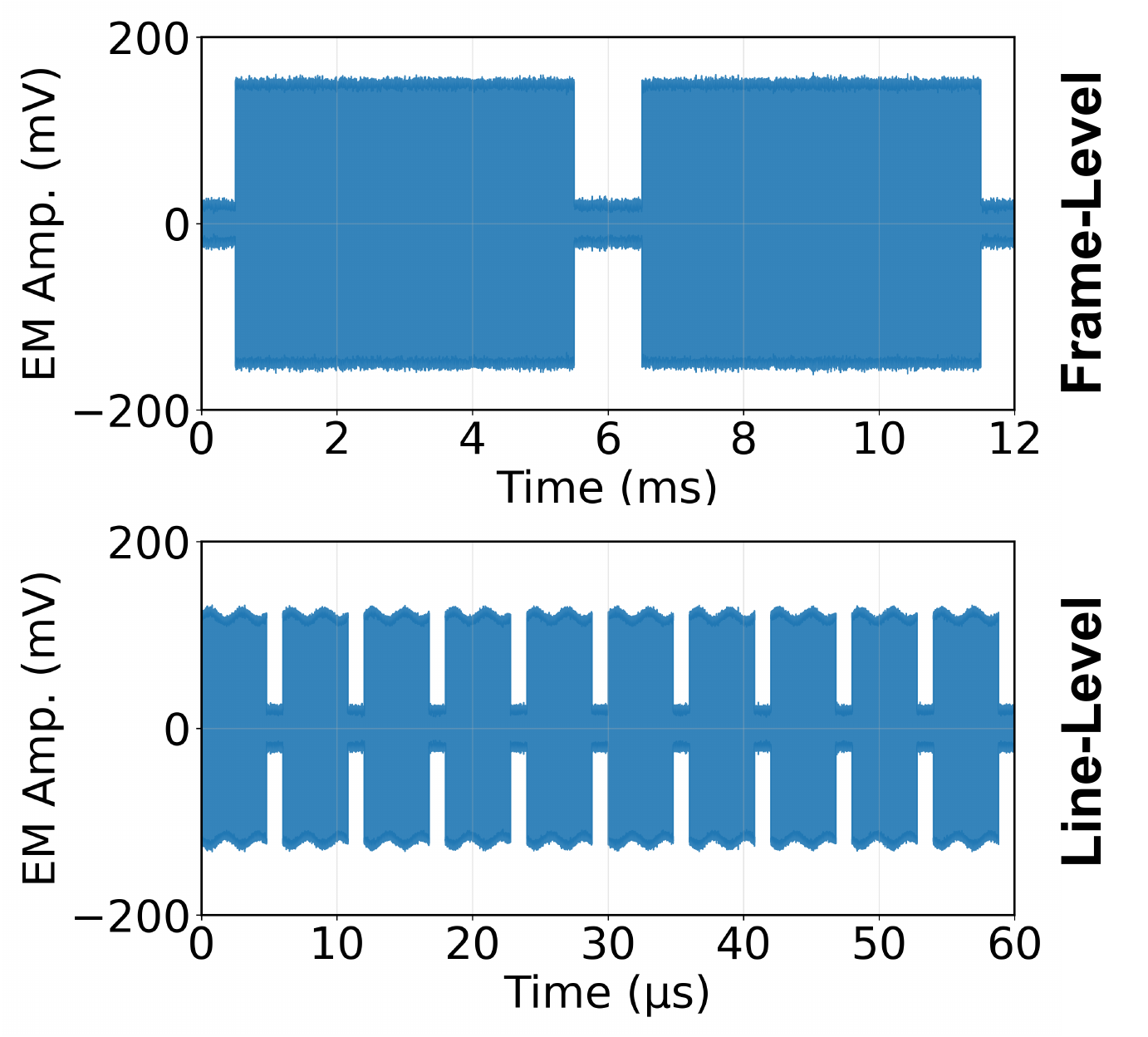}
        \caption{}
        \label{fig:sub_b}
    \end{subfigure}
    \vspace{-5pt}
    \caption{EM leakage in MIPI CSI-2 image transmission. (a) CSI-2 data organization. (b) Frame-level and Line-level transmission's EM leakage.}
    \label{fig:em_transmission}
    \vspace{-2em}
    \Description{This figure shows ...}
\end{figure}

\noindent\textbf{Information-bearing EM Emissions in MIPI CSI-2.} 
As illustrated in Figure~\ref{fig:sub_a}, CSI-2 organizes image transmission hierarchically~\cite{lee2021mipi}, with frames divided into rows and each row further decomposed into columns. Within each frame, the protocol structures the transmitted data into packets, specifically: each row transmission begins with a Line Start (LS) short packet, followed by a Long Packet containing a Header and Pixel Payload, and ends with a Line End (LE) short packet. Rows are separated by line blanking intervals, while frame blanking intervals delimit frame boundaries. This structured packetization not only enables reliable high-speed transmission but also induces distinctive EM emissions. As shown in Figure~\ref{fig:sub_b}, these emissions manifest on multiple time scales: at the frame level, aggregated signals appear as periodic bursts, each corresponding to one frame, whereas at the line level, finer-grained periodic patterns align with individual row transmissions.

%-------------------------------------------------------------------------------
\section{Threat Model}
% \subsection{Motivation of \app{}}
% Palm recognition has seen increasing adoption across both palmprint and palmvein modalities~\cite{grant2025palmretail,NGI,zhong2019centralized,ccc2023}. Compared to traits such as iris and face, it offers greater robustness to illumination, larger feature-rich regions, easier acquisition without user cooperation, and stronger resistance to spoofing, making it well-suited for practical deployment.

% Despite these advantages, palm recognition systems inevitably expose sensitive information during data transmission. Our research identifies a critical vulnerability: the EM leakage in this stage are strongly correlated with image data and can be reconstructed into raw images and subsequently restored into palm biometric patterns. Through denoising and enhancement, attackers can recover high-quality palmprint and vein features, posing severe risks since these immutable patterns cannot be revoked or replaced, leading to long-term privacy and security threats.

% To our best knowledge, no prior work has investigated EM side-channel leakage from palm recognition systems, and certainly none has addressed dual-mode palmprint–vein systems. Therefore, this study aims to explore the feasibility of such attacks and to assess their threat level to existing palm recognition deployments.

\begin{figure*}[t]
    \centering 
    \includegraphics[width=\linewidth]{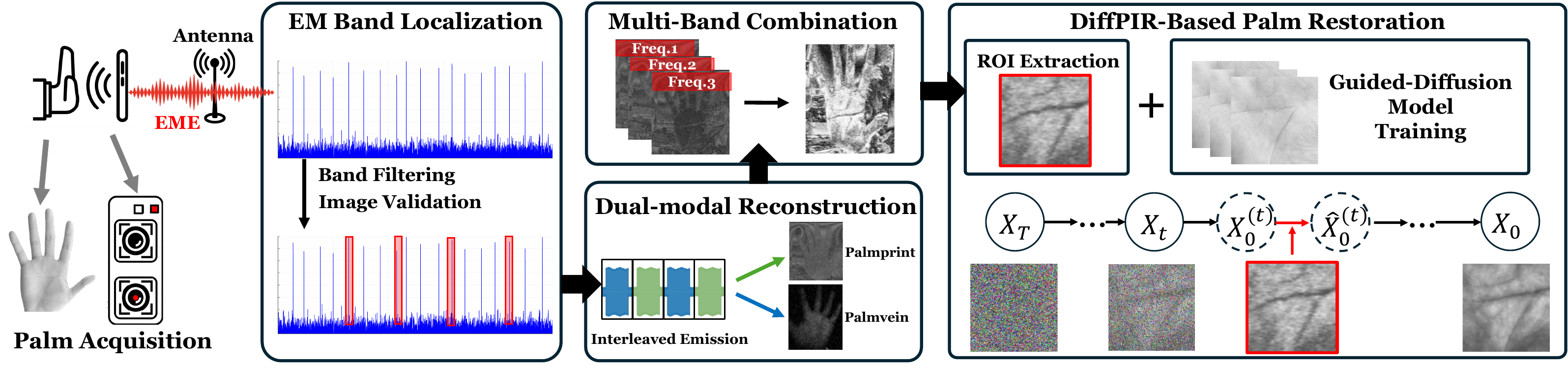}
    \vspace{-20pt}
    \caption{Overview of \app{}.}
    \vspace{-5pt}
    \label{fig:attack_design}
    \Description{This figure shows ...}
\end{figure*}

The adversary's objective is to exploit EM emissions leaked from biometric acquisition and recognition systems to reconstruct palm biometric features, thereby enabling unauthorized access, identity theft, and financial fraud.

\noindent\textbf{Victim Device.} The victim devices are biometric acquisition and recognition systems equipped with either single-mode or dual-mode cameras. During operation, raw data are transmitted via high-speed interfaces such as CSI2, which inevitably generate EM emissions that may expose sensitive biometric information.

\noindent\textbf{Adversary Capabilities.} The adversary cannot physically access or tamper with the victim systems, nor modify hardware, firmware, or software. However, by capturing the EM emissions leaked during image acquisition and real-time biometric recognition, the adversary can remotely extract data sufficient to recover palm biometric features. Using commercially available antennas, low-noise amplifiers (LNAs), and software-defined radios (SDRs), the adversary can operate from a concealed distance without raising suspicion.

\noindent\textbf{Attack Scenarios.} As shown in Figure~\ref{attack_scenario}, we consider real-world deployment scenarios where palm-based biometric systems are widely used, including secure building entry points, identity verification kiosks, and palm payment terminals deployed by major retailers~\cite{grant2025palmretail}. The eavesdropper discreetly installs compact EM signal capturing devices behind walls, under counters, or within fixtures near the target systems. When a user performs palm-related authentication, the concealed device proactively captures the EM emission leaked during the image acquisition process. The adversary is able to reconstruct a palm template just within a few seconds. 

\section{Attack Design}

Figure~\ref{fig:attack_design} provides an overview of \app{}. We first introduce its core four modules in terms of the overall workflow, and elaborate in the following respective subsections.

\textit{(1) EM Band Localization.} 
Palm-related emissions are embedded in a noisy wide spectrum, so this module identifies informative sub-bands carrying biometric information using a two-stage process: (i) statistical band filtering to discard noise-dominated regions, and (ii) image validation that reconstructs preliminary images to verify palm-relevant structures.

\textit{(2) Dual-Modal Image Reconstruction.} 
For each localized band, intercepted EM signals are transformed into palm images. While reconstruction is straightforward for single-modal systems, dual-modal systems are challenging due to asynchronously interleaved palmprint and palm vein transmissions. We design a disentanglement method to separate and align the two modalities, enabling synchronized dual-modal reconstruction.

\textit{(3) Multi-Band Combination.} 
Single-band reconstructions suffer from stochastic noise and bit-level ambiguities caused by bit-packed acquisition. To address this, we integrate reconstructions from multiple informative bands using a multi-band optimization strategy. By exploiting harmonic relationships across frequencies, this module consolidates complementary features, restores intensity variations, and preserves structural details.

\textit{(4) DiffPIR-Based Palm Restoration.} 
The fused images undergo ROI extraction and diffusion-based restoration. Building on DiffPIR~\cite{DDPIR}, we incorporate a \textit{structure-guided conditional prior} derived from degraded images to guide the diffusion process. This design suppresses heterogeneous noise, corrects band-limited distortions, and preserves key biometric micro-structures such as palmprint creases and vein bifurcations, producing reconstructions with high perceptual quality and \textbf{biometric faithfulness} suitable for spoofing attacks and downstream analysis.

\begin{figure}[t]
    \centering 
    \includegraphics[width=\columnwidth]{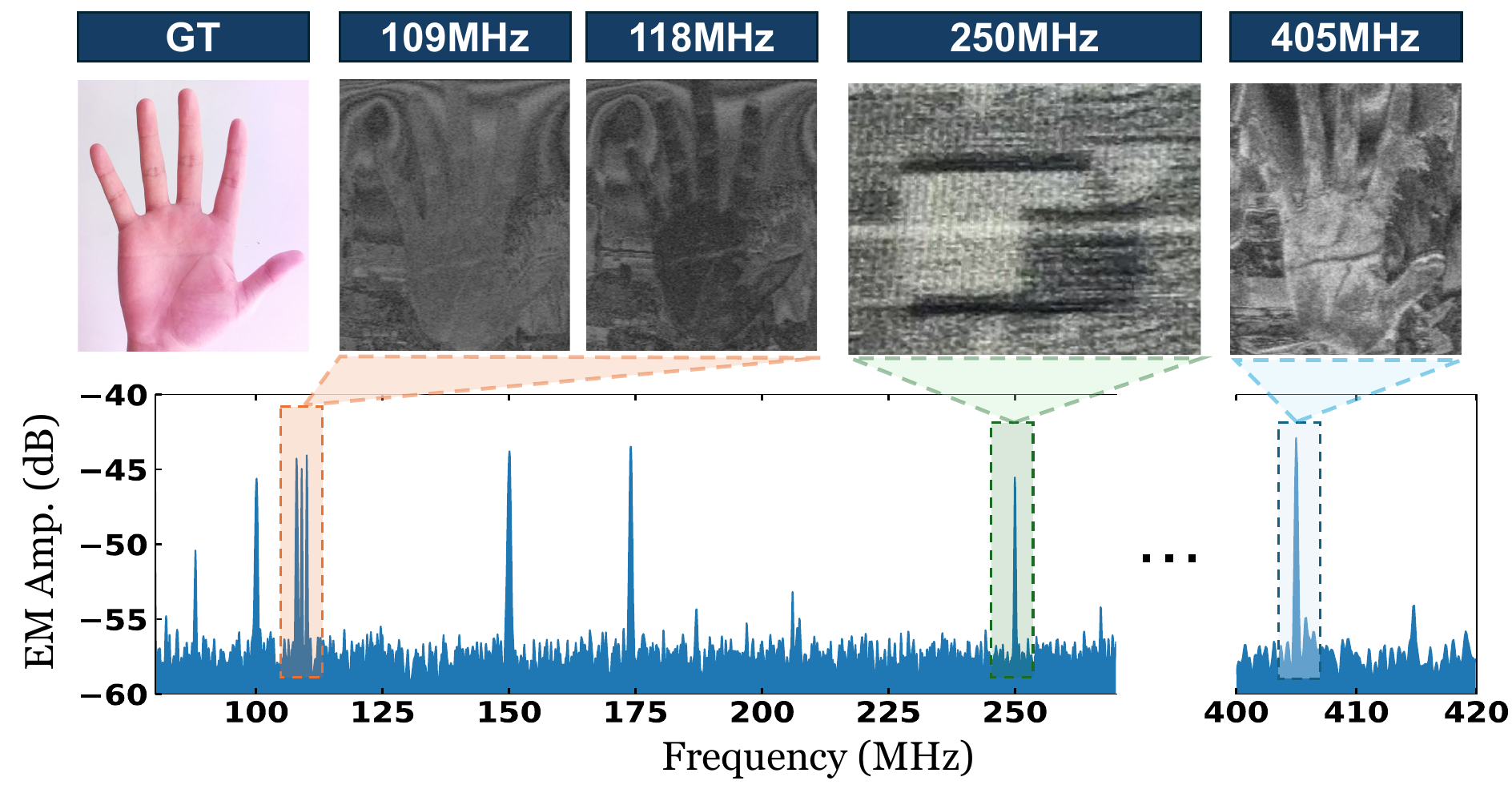}
    \vspace{-20pt}
    \caption{Illustration of signals from different frequencies.} 
    \label{fig:specturm}
    \vspace{-2em}
    \Description{This figure shows ...}
\end{figure}

\subsection{EM Leakage Bands Localization}
% --- Paragraph 1 ---
% To enable improved quality while reducing computational cost by avoiding uninformative or noisy spectral regions, we analyze the EM emissions generated by palm recognition systems. These emissions inherently have a wide frequency spectrum; however, only specific frequency sub-bands encapsulate meaningful biometric image content. 

Figure~\ref{fig:specturm} illustrates the diverse signal characteristics captured across different EM sub-bands. While certain frequencies—such as 109 MHz, 118 MHz, and 405 MHz—yield palm images with discernible biometric features, many other bands are dominated by irrelevant emissions or noise (e.g., the 250 MHz band reveals HDMI). Without prior knowledge, pinpointing a sub-band that contains useful biometric signals within a wide spectrum is a non-trivial task.

While each informative sub-band may capture only a partial and limited aspect of the palm’s structure, it can simultaneously exhibit strong structured noise patterns. This combination—limited signal coverage and dominant noise—amplifies the difficulty for downstream restoration, making it harder to recover a clean and complete biometric image from any single band. To address this, we aim to exhaustively identify signals from all sub-bands that may carry complementary biometric cues. 

This insight makes the problem significantly harder than single-band reconstruction: in practice, we do not know a priori how many informative bands exist or where they are located within the spectrum. To address this challenge, we propose an automated frequency identification method, outlined in Algorithm~\ref{alg:freq_select}, which integrates statistical signal characterization with visual interpretability. The method begins by partitioning the full EM spectrum $S(f)$ into discrete sub-bands over the range $[f_{\min}, f_{\max}]$ (Line~1), and proceeds in two stages: (1) \emph{Band Filtering}, where candidate bands are selected based on spectral energy and statistical features, and (2) \emph{Image Validation}, where preliminary reconstructions are assessed to confirm the presence of palm-relevant structures.

\begin{algorithm}[b]
\caption{Frequency Band Localization}
\label{alg:freq_select}
\KwIn{EM spectrum $S(f)$, frequency range $[f_{\min}, f_{\max}]$}
\KwOut{Informative sub-bands $\mathcal{F}_{\text{img}}$}

Divide $[f_{\min}, f_{\max}]$ into sub-bands $\{f_i\}_{i=1}^N$\;

\For{$i \gets 1$ \KwTo $N$}{
\tcp{Stage 1: Band Filtering}
    Extract $s_i(t)$ from $S(f_i)$\;
    \Indp
    $E_i = \|s_i(t)\|^2$ \tcp*{Signal energy} 
    $H_i = \mathcal{H}(\mathrm{FFT}(s_i(t)))$ \tcp*{Spectral ent.}  
    $A_i = \max(\mathrm{ACF}(s_i(t)))$ \tcp*{Autocorr. peak}  
    \Indm
\tcp{Stage 2: Image Validation}
    \If{$E_i > \theta_E$ \textbf{and} $A_i > \theta_A$ \textbf{and} $H_i < \theta_H$}{
        $I_i = \text{\textbf{\textsc{TempestSDR}}}{(f_i^{\text{low}}, f_i^{\text{high}})}$ \tcp*{SDR  Algo}
        \Indp
        $\mathcal{H}(I_i)$ \tcp*{Image entropy}  
        $\mathcal{E}(I_i) = \|\nabla I_i\|$ \tcp*{Edge intensity}  
        \Indm
        \If{$\mathcal{H}(I_i) > \theta_{\mathcal{H}}$ \textbf{and} $\mathcal{E}(I_i) > \theta_{\mathcal{E}}$}{
            $\mathcal{F}_{\text{img}} \gets \mathcal{F}_{\text{img}} \cup \{F_i\}$
        }
    }
}
\Return{$\mathcal{F}_{\text{img}}$}
\end{algorithm}

\noindent\textbf{Band Filtering (Lines 2--6).}  
For each sub-band, the time-domain signal $s_i(t)$ is extracted and evaluated using three metrics: energy $E_i$ (overall activity), spectral entropy $H_i$ (frequency regularity), and peak autocorrelation $A_i$ (temporal periodicity). Sub-bands with high $E_i$, low $H_i$, and strong $A_i$ are retained as structured, information-bearing candidates for further processing.  

\noindent\textbf{Image Validation (Lines 7--12).}  
Each candidate signal $s_i(t)$ is reconstructed into a grayscale image $I_i$ using \textsc{TempestSDR}~\cite{Marinov2014RemoteVE}:
\begin{equation}
P_{rec}^{[f_l,f_h]} = \mathcal{R}\{n(t) + b_{clk} + H_{[f_l,f_h]}[\mathcal{D}(P_{orig})]\},
\end{equation}
where $\mathcal{R}$ denotes the reconstruction operator and $H_{[f_l,f_h]}$ represents the EM transfer function.  
After reconstruction, two visual metrics are computed to ensure that each band captures palm-relevant structures rather than incidental artifacts: image entropy $\mathcal{H}(I_i)$, reflecting intensity diversity, and edge intensity $\mathcal{E}(I_i)$, emphasizing crease and vein patterns. While either metric alone may arise from noise, their joint prominence serves as a reliable indicator of palm-related content. Bands exhibiting high $\mathcal{H}(I_i)$ and $\mathcal{E}(I_i)$ values are retained as final candidates for subsequent processing.

\subsection{Dual-Modal Image Reconstruction}
\label{subsec: Reconstruction}

Although we utilize \textsc{TempestSDR} to reconstruct raw images to facilitate frequency localization, modern dual-mode palm recognition systems typically alternate between capturing palmprint and palmvein modalities~\cite{cadence2025csi2tx,lee2021mipi}. When \textsc{TempestSDR} is naively applied to such interleaved transmissions, the resulting reconstructions contain entangled content from both modalities, often mixed in unpredictable and non-uniform ways. As a result, these raw images are largely unusable for downstream processing, necessitating more sophisticated disentanglement strategies before any meaningful restoration or analysis can take place. 

To address this issue, we analyze the eavesdropped EM signals and observe that dual-modal systems follow specific transmission patterns. For synchronized frame-interleaved systems, palmprint and palmvein data alternate regularly across consecutive frames. We first detect the transmission mode by analyzing frame header signatures and inter-frame correlation patterns:

\begin{equation}
\rho_{inter} = \frac{1}{N-2}\sum_{k=1}^{N-2} \text{corr}(F_k, F_{k+2}),
\end{equation}
where $F_k$ represents the $k$-th frame. High $\rho_{inter}$ values (>0.8) indicate frame-alternating transmission, enabling temporal separation by frame parity:

\begin{equation}
M_k = \begin{cases}
k \bmod 2, & \text{if } \rho_{inter} > \tau \\
\text{Adaptive}. & \text{otherwise}
\end{cases}
\end{equation}

\noindent For systems with $\rho_{inter} > \tau$, we perform modality-specific reconstruction:

\begin{equation}
P_{print}[r,c] = \frac{1}{N_{print}} \sum_{j=0}^{N_{print}-1} |s_{IQ}^{(2j)}[r,c]|,
\end{equation}

\begin{equation}
P_{vein}[r,c] = \frac{1}{N_{vein}} \sum_{j=0}^{N_{vein}-1} |s_{IQ}^{(2j+1)}[r,c]|.
\end{equation}

However, real-world devices often exhibit asynchronous or line-interleaved transmissions due to sensor-level timing variations and SoC-specific architectures. For these cases ($\rho_{inter} \leq \tau$), we employ an adaptive synchronization mechanism that analyzes the vertical blanking interval patterns and horizontal synchronization signals embedded in the EM emissions. Specifically, we detect packet boundaries through spectral discontinuities in the baseband signal:

\begin{equation}
B_k = \arg\max_{t} \left| \frac{d}{dt} S(f_c, t) \right|,
\end{equation}
where $S(f_c, t)$ represents the signal power at carrier frequency $f_c$. These boundaries, combined with protocol-specific timing templates (e.g., MIPI CSI-2 packet headers), enable accurate modality classification even for non-uniform transmission patterns. The effectiveness of this adaptive approach ensures robust modality separation across diverse dual-modal architectures while maintaining compatibility with standard frame-alternating systems.

\subsection{Multi-band Image Combination}
\label{subsec:multi-band}
While the dual-modal image reconstruction effectively disentangles the modalities into separate palmprint and palmvein images, it inevitably incurs information loss due to the bit-packed acquisition formats commonly used in sensor hardware. In such formats, multiple bit positions are compressed into repeating binary patterns, which become electromagnetically indistinguishable within a single frequency band. This aliasing effect causes subtle grayscale variations to collapse, leading to noticeable gradient artifacts and the erosion of fine structural details in the reconstructed images.

Our key insight is that while individual frequency bands suffer from these ambiguities, the harmonic relationships across multiple bands preserve complementary information. When the fundamental frequency $f$ cannot differentiate between bit positions with identical periodicities, the harmonic at $2f$ often carries discriminative phase or amplitude variations necessary for accurate recovery. This observation motivates our multi-band optimization framework:  
\begin{equation}
    \min_{\alpha_i} \; \left\| S(I_{\text{reconstructed}}) - v_{\text{target}} \right\|^2 
    + \lambda \, \Phi(I_{\text{reconstructed}}),
\end{equation}
where the first term enforces intensity consistency over uniform regions, and $\Phi(\cdot)$ is a regularizer encouraging the preservation of structural details such as palm creases and vein edges.  

The reconstructed image is expressed as  
\begin{equation}
    I_{\text{reconstructed}} = \sum_{i=1}^{N} \alpha_i \cdot B_i(f_i^{\text{low}}, f_i^{\text{high}}),
\end{equation}
where $B_i$ denotes the filtered image obtained from frequency band $i$. The candidate bands are restricted to the validated outputs from the previous stage:
\begin{equation}
    \{B_i\}_{i=1}^N \subseteq \mathfrak{I}_{\text{img}},
\end{equation}
with $\mathfrak{I}_{\text{img}}$ denoting the set of informative sub-band reconstructions identified by the frequency localization algorithm.  

Here, $S(\cdot)$ denotes a segmentation operator for uniform regions, $v_{\text{target}}$ is their expected constant intensity, and the optimization adaptively assigns weights $\{\alpha_i\}$ to balance surface uniformity with preservation of palmprint and vein structures. In practice, amplitude thresholding suppresses noise before fusion, and the number of combined bands is selected to trade off reconstruction fidelity against computational cost.

\subsection{Diffusion-based Palm Restoration}
\label{subsec:Restoration}

While the proposed multi-band image combination alleviates bit-level grayscale collisions and restores critical structural details, practical EM side-channel acquisition of palmprint and palmvein still suffers from hardware mismatches, EM interference, and environmental noise. These factors introduce artifacts and distortions that obscure fine biometric details and reduce recognition quality.

\noindent\textbf{Problem Formulation.}
Following prior EM reconstruction works, we model the image restoration task as a linear inverse problem:
\begin{equation}
\mathbf{y} = \mathbf{H}\mathbf{x} + \mathbf{n},
\label{eq:linear_inverse}
\end{equation}
where $\mathbf{x} \in \mathbb{R}^n$ denotes the clean palm image, $\mathbf{y} \in \mathbb{R}^m$ the multi-band combined image (output of Section~\ref{subsec:multi-band}), $\mathbf{H} \in \mathbb{R}^{m\times n}$ the degradation operator, and $\mathbf{n} \sim \mathcal{N}(0, \sigma_y^2 \mathbf{I})$ additive Gaussian noise.

\noindent\textbf{Challenges in Palm EM Restoration.} Palm biometric restoration from EM signals introduces unique challenges. First, the degradation operator $\mathbf{H}$ is unknown and device-dependent, involving frequency-selective attenuation, phase distortions, and structured interference that vary across hardware configurations. Second, unlike supervised restoration methods that rely on paired degraded–clean samples, an adversary in a real-world side-channel attack cannot access the victim’s clean biometric images as training labels, since doing so would require compromising the biometric device itself and would contradict the stealthiness assumption of the attack.

\noindent\textbf{DiffPIR Framework for Plug-and-Play Restoration.}
To address these challenges, we adopt the plug-and-play DiffPIR framework~\cite{DDPIR}, which enables unsupervised restoration through alternating optimization. The framework solves the following optimization problem via Half-Quadratic Splitting (HQS):
\begin{equation}
\hat{\mathbf{x}} = \arg\min_{\mathbf{x}} \|\mathbf{y} - \mathbf{H}\mathbf{x}\|^2 + \lambda P(\mathbf{x}),
\end{equation}
where $P(\mathbf{x})$ represents a learned diffusion prior. When $\mathbf{H}$ is unknown or complex, DiffPIR assumes identity degradation ($\mathbf{H} \approx \mathbf{I}$) for pure denoising, aligning with our scenario where degradations stem primarily from additive EM interference~\cite{DDPIR}.

The framework alternates between two steps during inference:
\begin{align}
\text{(Prior):} \quad \mathbf{x}_0^{(t)} &= \frac{1}{\sqrt{\bar{\alpha}_t}}\left(\mathbf{x}_t + (1-\bar{\alpha}_t)\mathbf{s}_\theta(\mathbf{x}_t, t)\right), \label{eq:diffpir_denoise}\\
\text{(Data Fidelity):} \quad \hat{\mathbf{x}}_0^{(t)} &= \arg\min_{\mathbf{x}} \|\mathbf{y} - \mathbf{x}\|^2 + \rho_t\|\mathbf{x} - \mathbf{x}_0^{(t)}\|^2, \label{eq:diffpir_data}
\end{align}
where Eq.~\eqref{eq:diffpir_data} enforces consistency with the EM-reconstructed image $\mathbf{y}$, with $\rho_t = \lambda(\sigma_n/\bar{\sigma}_t)^2$ controlling data fidelity.

\noindent\textbf{Unsupervised Prior Learning.}
DiffPIR enables learning a powerful diffusion prior $\mathbf{s}_\theta$ from only publicly available clean palm datasets~\cite{CASIA,zhang2017towards,hao2008multispectral,luo2024palm}, avoiding the need for any paired EM–clean data that would violate the stealthiness constraint of our threat model. The prior captures the manifold of palmprint ridges and vein structures via denoising-score matching~\cite{DDRM,ho2020denoising}.

\noindent\textbf{Structure-Guided Conditioning.}
To prevent hallucinated ridge patterns and ensure semantic consistency, we condition the denoiser on the multi-band combined EM reconstruction $\mathbf{y}$ itself. Despite noise, $\mathbf{y}$ retains coarse ridge flow and palm topology, which anchors the restoration to physically leaked biometrics rather than free-form generative priors:
\begin{equation}
\mathbf{x}^{(t)}_0=
\frac{1}{\sqrt{\bar{\alpha}_t}}\!\left(\mathbf{x}_t+
(1-\bar{\alpha}_t)\,\mathbf{s}_\theta(\mathbf{x}_t,t,\mathbf{y})\right).
\end{equation}
This lightweight guidance requires no additional feature engineering nor domain-specific annotations.

\noindent\textbf{Preventing Generative Hallucination.}
We preserve EM-grounded identity information through dual constraints: (i) The data-fidelity term in Eq.~\eqref{eq:diffpir_data} anchors each reverse diffusion step to the observed EM leakage $\mathbf{y}$, with adaptive weight $\rho_t$ maintaining strong coupling throughout denoising. (ii) Structure-guided conditioning directly injects $\mathbf{y}$ into the denoiser network, ensuring generated patterns remain consistent with physical EM emanations. This optimization-network dual constraint ensures restored biometric features originate from actual EM leakage rather than learned priors.

\section{Evaluation}

To comprehensively assess the effectiveness of \app{}, we conduct a three-stage evaluation across diverse hardware platforms and real-world scenarios involving 25 human participants.
\textbf{\textit{First}}, we evaluate the image restoration capability, examining how accurately \app{} can recover palmprint and palmvein images from intercepted EM signals.
\textbf{\textit{Second}}, we assess spoofing effectiveness by testing whether the reconstructed images can successfully deceive state-of-the-art palm recognition models.
\textbf{\textit{Finally}}, we examine the robustness of \app{} under varying environmental and operational conditions to validate its practical feasibility.

% We conduct comprehensive experiments of \app{} on diverse hardware and real-world scenarios involving ten human participants, to validate its practical feasibility in both recovery fidelity and spoofing attack effectiveness. 

% We start with introducing the experimental setup in Section \ref{subsec:exp_setup}, then evaluate the effectiveness of \app{} in Section~\ref{subsec:EFFECTIVE}, and finally examine the impact of practical factors on \app{} in Section~\ref{subsec:impact}.

\subsection{Experimental Setup}
\label{subsec:exp_setup}

\textbf{Hardware.} To reproduce palm recognition processes, we built a modular acquisition platform using single-board computers (SBCs) connected to visible-light and Near-Infrared sensors. The SBC controllers include Raspberry Pi 3B+ (S1), Raspberry Pi 5 (S2) and NVIDIA Jetson Nano (S3). We use three devices for palmprint acquisition: OV5647 (V1), IMX219 (V2) and IMX708 (V3), and use three NIR devices for palmvein acquisition: 23H166-LED (IR1), IMX219-160 (IR2) and HW200 (IR3). Besides the above single modal devices, a dual-modal device, HAOKAI-H220 (DUAL), is employed for simultaneous palmprint and palmvein capture. To further evaluate \app{}’s performance against real-world devices, we include three commercial off-the-shelf (COTS) devices C1, C2, and C3. We withhold disclosure of the exact models of the tested commercial devices to provide vendors time to develop solutions addressing risks. 

% Table \ref{tab:devices} summarizes the palm acquisition devices.

\begin{figure}[t]
    \centering 
    \includegraphics[width=\linewidth]{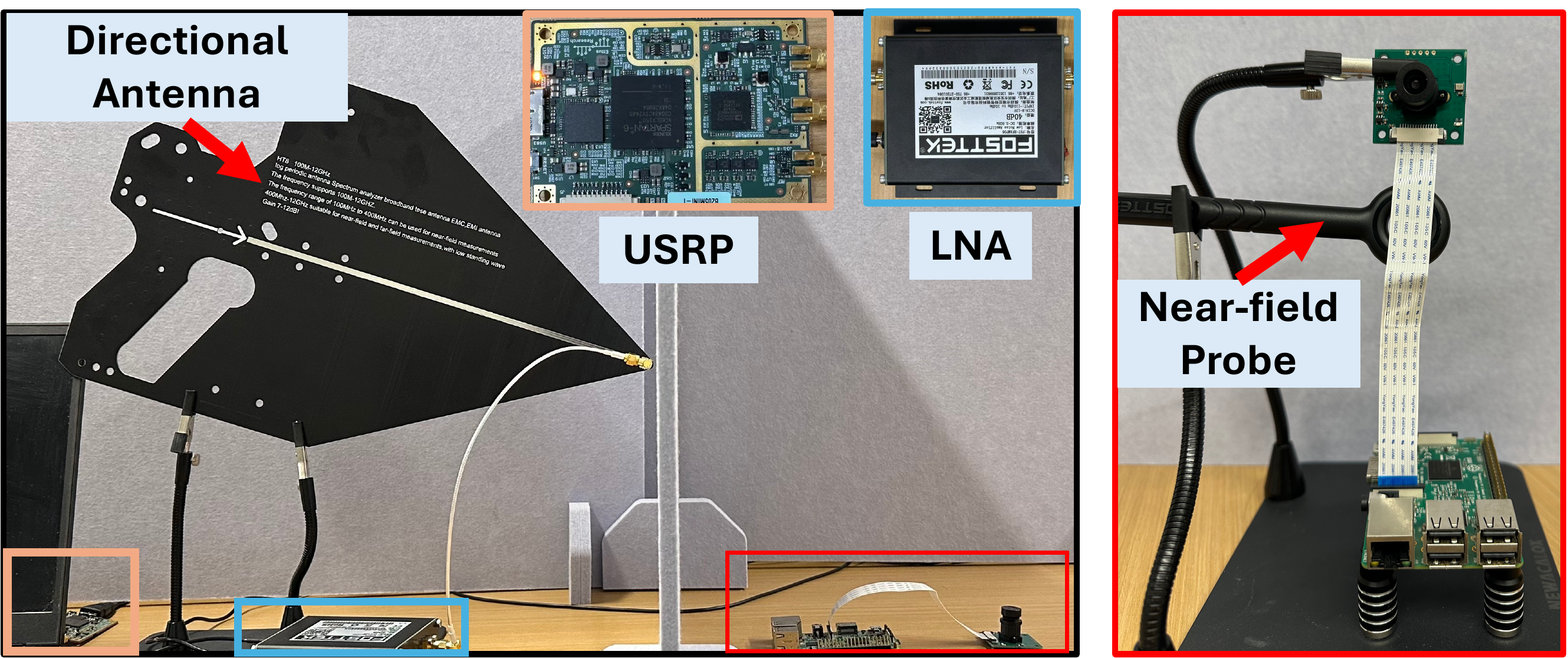}
    \caption{EM signals acquired using a directional antenna and a near-field probe.} 
    \label{fig:expsetup}
    \vspace{-1.5em}
    \Description{This figure shows ...}
\end{figure}

Figure~\ref{fig:expsetup} illustrates the EM acquisition system, which is built on a Universal Software Radio Peripheral (USRP) B200 SDR~\cite{ettus2023usrp}, equipped with a FOSTTEK near-field magnetic probe for close-range measurements or an Eujgoov directional antenna (0.1–12 GHz) for long-range reception. We use a FOSTTEK FST-RFAMP06 low-noise amplifier (LNA) with a gain of 40 dB to enhance weak EM emissions. The USRP operates at a sampling rate of 10 MS/s with an RF bandwidth of 20 MHz.

% \begin{table}[t]
% \centering
% \caption{Palm Acquisition Devices. }
% \label{tab:devices}
% \resizebox{0.80\columnwidth}{!}{
% \begin{tabular}{l l c }
% \toprule
%  \textbf{Type} & \textbf{Device} & \textbf{Task} \\\hline
% \multirow{7}{*}{Prototype
% }&OV5647 (V1) & print  \\
% &IMX219 (V2) & print  \\
% &IMX708 (V3) & print \\
% & 23H166-LED (IR1) & vein \\
% &IMX219-160 (IR2) & vein \\
% &HW200 (IR3)& vein \\
% & HAOKAI-H220(DUAL) & dual  \\\midrule
% \multirow{2}{*}{Commercial} & C1 & dual \\
% & C2 & dual  \\
% \bottomrule
% \end{tabular}}
% \end{table}

\noindent\textbf{Software.} For the configuration of USRP, we employ TempestSDR on the Ubuntu(24.04.5). For Diffusion training, as described in Section~\ref{subsec:Restoration}, we use PyTorch (2.4.0) with CUDA ( 12.1).

\noindent\textbf{Physical Deployment.} As shown in Figure~\ref{fig:expsetup}, to evaluate \app{}, we setup the attack against the target palm recognition system in both close-range and long-range configurations. In the close-range setting, a magnetic field probe is positioned near the transmission interface between the image sensor and the SBC with minimal interference, and in the long-range setting, a directional antenna intercepts radiated emissions without physical contact. 

% The captured EM waveforms exhibit fine-grained modulation patterns that can be reconstructed into preliminary biometric images directly reflecting the transmitted bitstream, and these reconstructions are further enhanced through our restoration pipeline, ultimately yielding high-quality palm biometric images for spoofing and advanced analysis.

% \noindent\textbf{Datasets.} We employ four public benchmark datasets of two palmprint and two palmvein datasets for restoration diffusion model training, including the SCUT-PV-v1 (SCUT) dataset~\cite{luo2024palm,ma2023focal,kang2014contactless}, the CASIA Multi-Spectral Palmprint Image Database V1.0 (CASIA-M)~\cite{hao2008multispectral}, the Tongji Contactless Palmprint Dataset (Tongji)~\cite{zhang2017towards} and  the CASIA Palmprint Image Database (CASIA)~\cite{CASIA}. To ensure meaningful evaluation on high-quality target models, we merge the Tongji and CASIA datasets into a combined palmprint dataset, and merge the SCUT and CASIA-M into a combined palmvein dataset.  Table \ref{tab:datasets} summarizes the dataset statistics.

\noindent\textbf{Diffusion Models for Restoration.} To account for modality differences, we train two separate diffusion models for palmprint and palmvein restoration. Table~\ref{tab:datasets} summarizes the dataset statistics. For palmprint, we train on the combined Tongji~\cite{zhang2017towards} and CASIA~\cite{CASIA} datasets; for palmvein, we use the combined SCUT~\cite{luo2024palm} and CASIA-M~\cite{hao2008multispectral} datasets. To prevent any identity leakage between generative and discriminative training stages, each combined dataset is partitioned at the \emph{subject level} (600 for palmprint and 650 for palmvein) into two disjoint halves: 50\% of subjects are exclusively used for diffusion model training, while the remaining 50\% are reserved for training the target recognition models. 
 Critically, our 25 test volunteers are not included in these public datasets, eliminating training data leakage. Once trained, each diffusion model is applied to the eavesdropped EM measurements collected from victim interactions: we feed the intercepted signals through the corresponding modality model to reconstruct palm images. These reconstructed images constitute the stolen biometric data and are subsequently used as spoofing probes against target recognition systems.

\begin{table}[h]
\centering
\vspace{-1em}
\caption{Dataset statistics and partition strategy for diffusion and recognition model training.}
\vspace{-1em}
\label{tab:datasets}
\resizebox{\columnwidth}{!}{
\begin{tabular}{l l c c c c}
\toprule
\textbf{Dataset} & \textbf{Task} & \textbf{\# Image} & \textbf{\# Subject} & \textbf{Diffusion} & \textbf{Recognition} \\\hline
SCUT & vein & 11,000 & 550 & 275 (50\%) & 275 (50\%) \\
CASIA-M & vein & 7,200 & 100 & 50 (50\%) & 50 (50\%) \\
Tongji & print & 12,000 & 300 & 150 (50\%) & 150 (50\%) \\
CASIA & print & 5,502 & 300 & 150 (50\%) & 150 (50\%) \\\midrule
CASIA + Tongji & print & 17,502 & 600 & 300 (50\%) & 300 (50\%) \\
CASIA-M + SCUT & vein & 18,200 & 650 & 325 (50\%) & 325 (50\%) \\
\bottomrule
\vspace{-1em}
\end{tabular}}

\raggedright\footnotesize{\textit{Note: For the CASIA dataset, 12 subjects (out of 312) were excluded due to incomplete data samples, resulting in 300 utilized subjects used in our settings.}}
\vspace{-5pt}
\end{table}

\noindent\textbf{Target Palm Recognition Models for Spoofing. }We evaluate our spoofing attack against two categories of target palm recognition models: palmprint-based and palmvein-based.
For palmprint-based models, we follow PCE-Palm~\cite{jin2024pce} and Diff-Palm~\cite{jin2025diff}, adopting three backbones, ResNet50~\cite{He2015}, MobileFaceNet~\cite{chen2018mobilefacenets}, and PalmNet~\cite{genovese2019palmnet}, with an input size of 224$\times$224, all trained using ArcFace~\cite{deng2019arcface} (margin $m$=0.5, scale $s$=48).
For palmvein-based models, we follow PVTree~\cite{shang2025pvtree} and adopt ResNet101~\cite{He2015} trained with ArcFace ($m$=0.5, $s$=64) for 20 epochs on real datasets.
Table~\ref{tab:target_models} summarizes all target models, their training datasets, and true accept rates.
Spoofing is evaluated in a 1:100 identification setting: reconstructed and diffusion-restored images are directly used as input probes against enrolled galleries, and we measure whether the target models accept these probes as genuine.

\begin{table}[h]
\centering
\caption{Target recognition models for attack evaluation.}
\vspace{-5pt}
\label{tab:target_models}
\resizebox{1.0\linewidth}{!}{
\begin{tabular}{l l c c}
\hline
\textbf{Model} & \textbf{Task} & \textbf{Training Dataset} & \textbf{TAR@
1e-4} (\%)\\
\hline
ResNet50~\cite{He2015} & Print  &  50\% CASIA  + Tongji & 94.81 \\
MobileFaceNet~\cite{chen2018mobilefacenets} & Print  & 50\% CASIA + Tongji & 96.26\\
PalmNet~\cite{genovese2019palmnet} & Print & 50\% CASIA + Tongji & 93.80\\
ResNet101~\cite{He2015} & Vein & 50\% CASIA-M + SCUT & 94.87\\
\hline
\end{tabular}}
\end{table}

\begin{table*}[t]
\centering
\caption{Comparison of reconstruction quality across palmprint, palmvein, and dual-modal settings among EMIRIS, EMEye, and our \app{}. A dash (—) indicates that the baseline method could not reconstruct dual-modal images and spoof models under the tested configuration.}
\vspace{-1em}
\label{tab:modality_comparison}
\resizebox{\textwidth}{!}{
\begin{tabular}{l|cccc|cccc|cccc}
\toprule
\multirow{2}{*}{\textbf{Method}} &
\multicolumn{4}{c|}{\textbf{Palmprint}} &
\multicolumn{4}{c|}{\textbf{Palmvein}} &
\multicolumn{4}{c}{\textbf{Dual-Modal}} \\
\cmidrule(lr){2-5} \cmidrule(lr){6-9} \cmidrule(lr){10-13}
 & SSIM $\uparrow$ & PSNR (dB) $\uparrow$ & FID $\downarrow$ & SSR (\%) $\uparrow$
 & SSIM $\uparrow$ & PSNR (dB) $\uparrow$ & FID $\downarrow$ & SSR (\%) $\uparrow$
 & SSIM $\uparrow$ & PSNR (dB) $\uparrow$ & FID $\downarrow$ & SSR (\%) $\uparrow$\\
\midrule
EMEye & 0.51 & 18.0 & 26.3 & --- & 0.42 & 18.7 & 29.4 & --- & --- & --- & --- & --- \\
EMIRIS  & --- & --- & --- & --- & 0.49 & 22.0 & 16.3 & 52.83 & --- & --- & --- & --- \\
\textbf{\app{} (Ours)} & \textbf{0.71} & \textbf{28.3} & \textbf{7.7} & \textbf{68.93} & \textbf{0.67} & \textbf{23.9} & \textbf{8.2} & \textbf{66.01} & \textbf{0.68} & \textbf{24.1} & \textbf{8.7} & \textbf{66.51}\\
\bottomrule
\end{tabular}
}
\end{table*}

\begin{figure*}[t]
    \centering 
    \includegraphics[width=\linewidth]{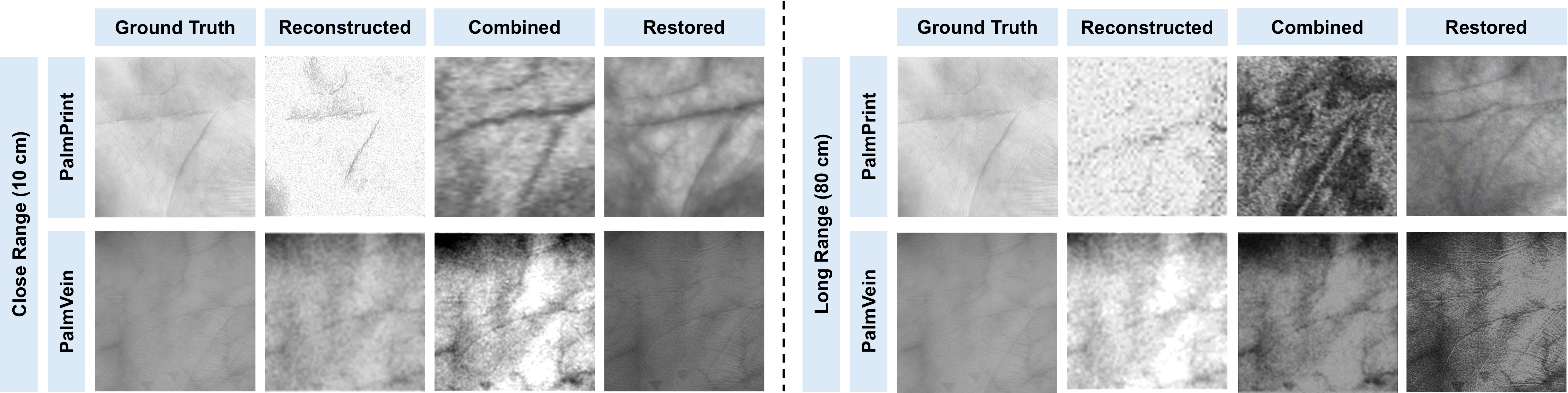}
    \vspace{-15pt}
    \caption{Reconstruction real-time human users, showing examples of palmprint (device V1, random select) and palmvein (device IR1, random select) in the single-modal setting, with close-range acquisition of live subjects data at 10 cm (left) and long-range acquisition at 80 cm (right). Ground Truth: the original high-quality palm print image; Reconstructed: the initial single-band reconstructed image; Combined: the image obtained by fusing reconstructed images from multiple frequency bands; Restored: the image restored from the combined image by diffusion model. } 
    \label{fig:printeva}
    \vspace{-1em}
    \Description{This figure shows ...}
\end{figure*}

% \noindent\textbf{Restoration Model Training. }We employ DiffPIR as our image restoration model. To account for modality differences, we train two separate models for palmprint and palmvein with their respective dataset. Following~\cite{DDPIR}, the network is configured with $64$ channels, two residual blocks per resolution, and attention at the $16\times16$ resolution. We set the diffusion process to $1{,}000$ steps with a linear noise schedule, and adopt the Adam optimizer with a learning rate of $2\times 10^{-4}$.

\noindent\textbf{Evaluation Metrics.} To ensure objective and domain-aligned assessment, we adopt standard evaluation metrics widely used in the literature on biometric spoofing and side-channel attacks~\cite{emiris,long2024eye}.

\begin{itemize}
\item \textit{Peak Signal-to-Noise Ratio (PSNR):} Evaluates pixel-wise fidelity between reconstructed and ground truth images, higher values indicate better pixel-level reconstruction accuracy.

\item \textit{Structural Similarity Index Measure (SSIM):} Assesses perceptual similarity in terms of luminance, contrast, and structure, ranging from -1 to 1, where 1 indicates perfect similarity.

\item \textit{Fréchet Inception Distance (FID):} Measures perceptual quality by comparing deep feature statistics, lower values indicate reconstructed images are closer to real ones in feature space.

\item \textit{Spoof Success Rate (SSR):} Quantifies the proportion of reconstructed palmprint and palmvein images that successfully bypass target biometric recognition models. A higher SSR indicates greater susceptibility of the recognition system to EM side-channel-based spoofing attacks.
\end{itemize}

Among these metrics, PSNR, SSIM, and FID evaluate the visual reconstruction quality of restored images, while SSR directly measures the attack effectiveness by assessing whether reconstructed biometric samples can successfully deceive recognition systems.

\subsection{Effectiveness Evaluation}
\label{subsec:EFFECTIVE}
% \textbf{Effectiveness of Multi-band Combination. }We evaluate the effectiveness of our multi-band combination strategy. As discussed in Section~\ref{subsec:multi-band}, single-band restoration often suffers from gradient artifacts on uniform regions due to the information loss in data acquisition. To address this issue, we integrate multiple frequency bands, leveraging their harmonic relationships to recover complementary information that is missing in any individual band. Figure~\ref{fig:experiments_metirc1} shows that the multi-band combination achieves consistently higher reconstruction quality, with an average SSIM of 0.70, FID of 10.81, and PSNR of 26.1 dB, compared to the single-band restoration which yields an SSIM of 0.25, FID of 21.28, and PSNR of 20.5 dB. The superior metrics demonstrate that multi-band fusion effectively reduces gradient artifacts, preserves structural details, and enhances the fidelity of palmprint and palmvein reconstruction.

We progressively evaluate \app{} across multiple dimensions, including its effectiveness in single and dual modal restoration, its ability to spoof target recognition models, and its performance in attacking real-world COTS devices. All experiments are conducted while 25 users operate the devices in real time, each user performed 10 interaction trials, yielding a total of 250 captured images.

% \noindent\textbf{Restoration Quality Comparison. }We reproduced the EMEye and EMIRIS methods under our palm acquisition setup and evaluated them using the same testing protocol. As shown in Table~\ref{tab:modality_comparison}, both baseline methods are inherently limited to single-stream processing and thus cannot handle dual-modal reconstruction. EMEye failed to produce spoofing-capable outputs because it lacks a restoration stage and does not employ a diffusion-based generative prior, resulting in blurry and low-fidelity reconstructions. EMIRIS, while somewhat effective on palmvein images, was also restricted to single-modality operation—it was originally designed for NIR-based iris sensing and cannot generalize to visible-spectrum palmprint or mixed dual-modal inputs. In contrast, \app{} delivers substantially higher reconstruction quality and achieves the highest spoofing success rates across all modalities.

\noindent\textbf{Restoration Quality Comparison.}
We evaluated the restoration quality of \app{} by comparing it with two representative EM-based biometric reconstruction methods, EMEye~\cite{long2024eye} and EMIRIS~\cite{emiris}. Although neither work was designed for palm restoration, both share conceptual similarities with our setting in that they exploit EM side-channel leakage to recover visual biometric information. Specifically, EMEye targets EM-based video frame inference, whereas EMIRIS reconstructs iris textures from NIR-driven EM emissions. Despite focusing on different biometric modalities, both exemplify EM-to-image recovery and thus offer meaningful baselines for evaluating the challenges of palm reconstruction.

For fair comparison, we reproduced the EMEye and EMIRIS pipelines under our palm acquisition setup and evaluated them using the same testing protocol. As shown in Table~\ref{tab:modality_comparison}, both baseline methods are inherently limited to single-stream processing and therefore cannot support dual-modal palm reconstruction. EMEye fails to produce spoofing-capable outputs because it lacks a dedicated restoration stage and does not incorporate a diffusion-based generative prior, which results in blurry, distorted, and low fidelity reconstructions. EMIRIS performs somewhat better on palmvein images but remains restricted to single modality operation, as it was originally designed for NIR-only iris sensing and cannot generalize to visible spectrum palmprint data or mixed dual-modal inputs. In contrast, \app{} delivers substantially higher reconstruction quality across both modalities and achieves the highest spoofing success rates in all evaluation settings.

\noindent\textbf{Effectiveness of Single-Modal Restoration.}  We first evaluated \app{} on single-modal restoration using three palmprint (V1-V3) and three palmvein (IR1-IR3) devices. Figure~\ref{fig:printeva} presents all intermediate and final images recovered by each stage of the \app{}, under both close and long-range settings. As shown, \app{} progressively refines the image through each stage, ultimately producing restored images that closely approximate the ground truth. 

\begin{figure}[t]
    \centering
    \subcaptionbox{Palmprint \label{fig:single_palm}}{
\includegraphics[width=0.95\linewidth]{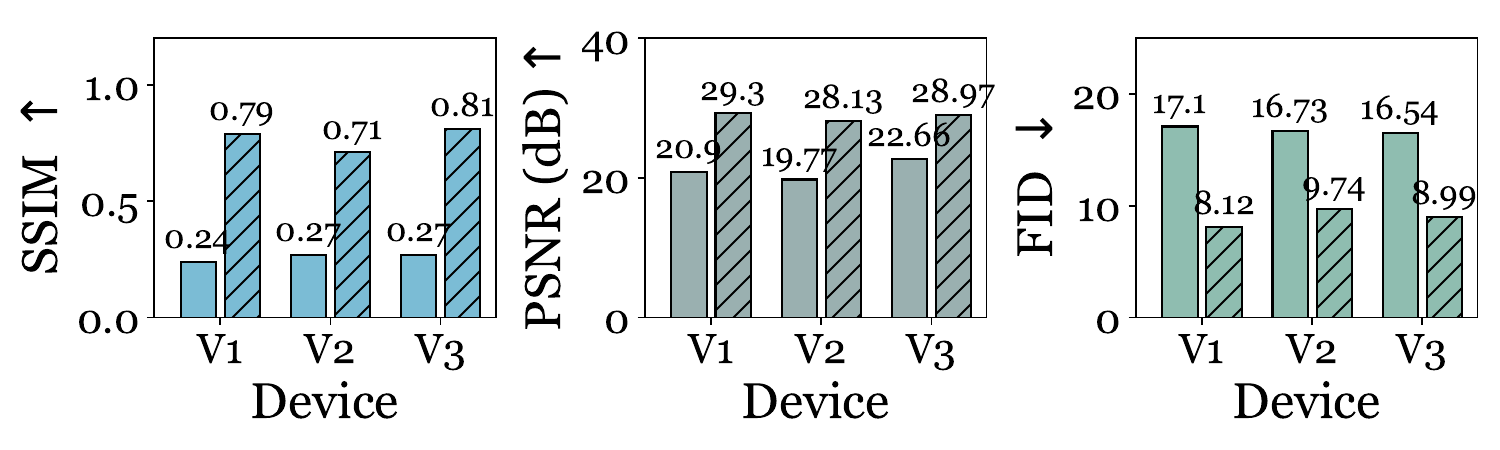}

    }
    \subcaptionbox{Palmvein\label{fig:single_vein}}{\includegraphics[width=0.95\linewidth]{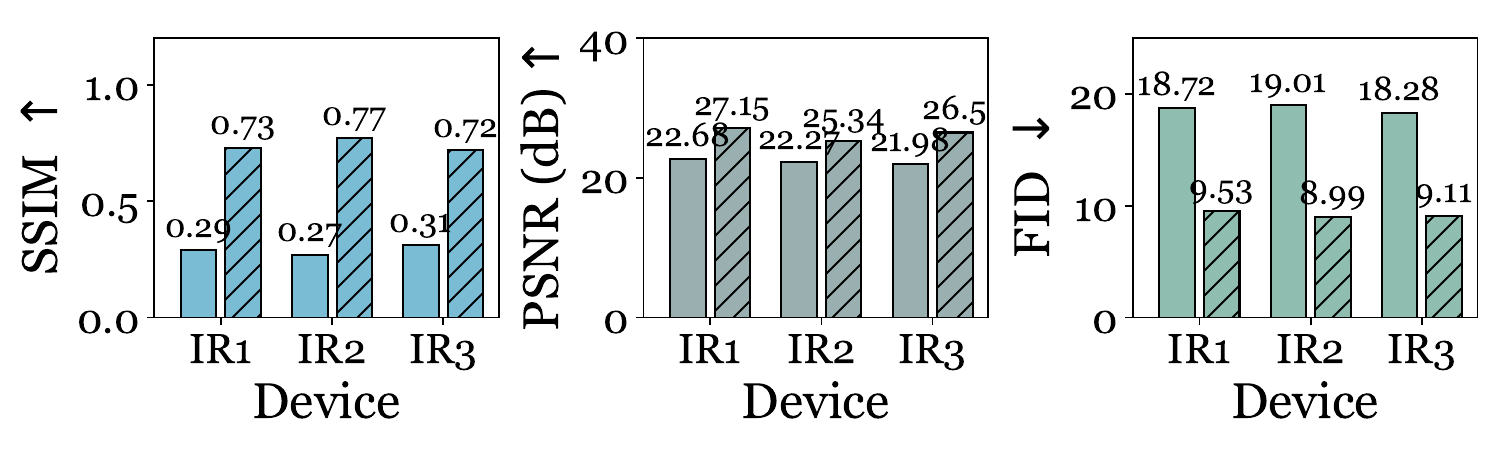}
    }
    \vspace{-1em}
    \caption{SSIM, PSNR and FID of \app{} on Single Modal. Solid bars: single band, hatched bars: multi-bands combined. }
    \vspace{-1em}
    \label{fig:exp_single_metric}
    \vspace{-1em}
    \Description{This figure shows ...}
\end{figure}

Figure~\ref{fig:exp_single_metric} further reports the quantitative metrics (SSIM, PSNR, and FID) across all devices, comparing \app{} with and without the proposed multi-band combination (hatched vs. solid bars). The solid bars represent single-band restoration, while the hatched bars indicate our multi-band fusion results. On palmprint devices, \app{} with multi-band fusion achieves up to 0.81 SSIM, 29.3 dB PSNR, and 8.12 FID; on palmvein devices, it achieves up to 0.77 SSIM, 27.15 dB PSNR, and 8.99 FID. The slightly lower metrics on palmvein reflect its inherent stability and robustness against external perturbations, making reconstruction more challenging. Nevertheless, \app{} still extracts high-fidelity representations across both modalities, demonstrating strong generalizability.

Comparing the two variants, multi-band combination yields consistent and significant gains across all metrics, confirming our hypothesis in \textbf{Section~\ref{subsec:multi-band}}. Specifically, SSIM increases by 0.55 (palmprint) and 0.50 (palmvein), PSNR by 8.4 dB (palmprint) and 4.52 dB (palmvein), while FID decreases by 8.98 (palmprint) and 10.02 (palmvein), confirming that multi-band combination improves reconstruction quality in structural and perceptual dimensions.

\begin{figure}[t]
    \centering
    \subcaptionbox{ Palmprint\label{fig:dual_palm}}{
\includegraphics[width=0.95\linewidth]{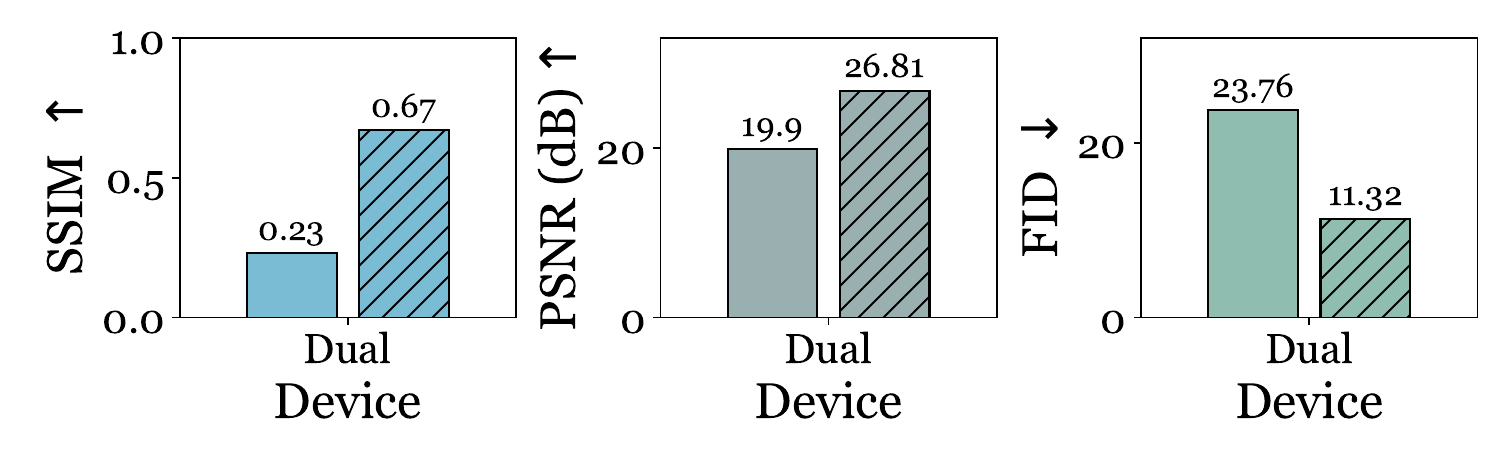}
    }
    \subcaptionbox{ Palmvein\label{fig:dual_vein}}{\includegraphics[width=0.95\linewidth]{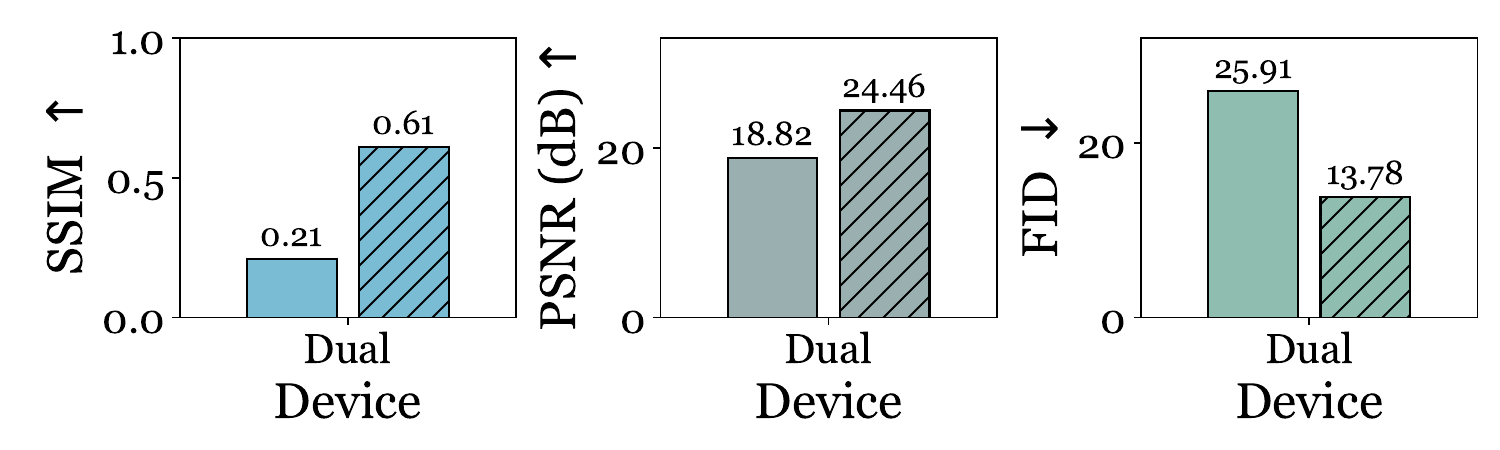}
    }
    \vspace{-1em}
    \caption{SSIM, PSNR and FID of \app{} on Dual Modal. Solid bars: single band, hatched bars: multi-bands combined.}
    \vspace{-1em}
    \label{fig:exp_dual_metric}
    \vspace{-1em}
    \Description{This figure shows ...}
\end{figure}

\noindent\textbf{Effectiveness of Dual-Modal Restoration.}
Building upon the single-modal results, we next evaluate \app{} under the dual-modal acquisition setting, where both palm-print (visible) and palm-vein (NIR) signals are captured simultaneously within a single sensing process. \app{} separates these interleaved data streams and reconstructs each modality independently from the same EM capture. Figure~\ref{fig:exp_dual_metric} summarizes the quantitative performance. From jointly acquired data, \app{} achieves 0.67 SSIM, 26.81 dB PSNR, and 11.32 FID on the palmprint modality, and 0.61 SSIM, 24.46 dB PSNR, and 13.78 FID on the palmvein modality. These results confirm that the proposed signal-separation and reconstruction framework can effectively disentangle and restore both biometric modalities from a single EM observation (\textbf{Section~\ref{subsec: Reconstruction}}).

Compared with the single-band variant (solid), incorporating multi-band combination (hatched) continues to yield substantial improvements, even under the intertwined dual-stream condition. For palmprint, SSIM increases by 0.40, PSNR by 5.64 dB, and FID decreases by 12.13; for palmvein, SSIM improves by 0.44, PSNR by 6.91 dB, and FID decreases by 12.44. These findings demonstrate that multi-band combination remains crucial for high-fidelity reconstruction when recovering two concurrently transmitted biometric channels from the same acquisition session.

\subsection{Effectiveness of Spoofing Target Models. }
To ensure a fair and representative evaluation, we follow the prior palm recognition works~\cite{shang2025pvtree,jin2025diff,jin2024pce}, which introduce advanced generative or enhancement pipelines for producing high-quality palm datasets and recognition benchmarks. These works have established strong CNN-based architectures validated on large-scale palm datasets, forming a solid and widely adopted foundation for subsequent research. We therefore adopt their recognition models, as summarized in Table~\ref{tab:target_models}, to provide consistent and credible baselines for assessing the spoofing effectiveness of \app{}.

% To ensure a fair evaluation of our attack,  we follow the prior palm recognition works~\cite{shang2025pvtree,jin2025diff,jin2024pce} and adopt the recognition models described in Table~\ref{tab:target_models}. These architectures represent widely used and well-validated CNN baselines, providing a representative foundation for our evaluation. 

\begin{table}[h]
\centering
\caption{Attack Success Rate (SSR) against different palm recognition models (mean $\pm$ std).}
\vspace{-10pt}
\label{tab:ssr_results}
\begin{tabular}{ccccc}
\toprule
\textbf{Model} & ResNet50 & MobileFaceNet & PalmNet & ResNet101 \\
\midrule
\textbf{SSR (\%)} & 
$68.0 \pm 2.3$ & 
$62.1 \pm 2.0$ & 
$70.0 \pm 2.5$ & 
$61.3 \pm 1.9$ \\
\bottomrule
\end{tabular}
\end{table}
Table~\ref{tab:ssr_results} presents the spoofing success rates (SSR) achieved by \app{} against different target models. The results demonstrate substantial effectiveness across all tested architectures, with an overall average spoofing success rate of 65.3\%. Among the palmprint models, PalmNet (print) exhibits the highest vulnerability with success rates reaching approximately 72\%, while ResNet50 (print) achieves around 68\% and MobileFaceNet (print) shows slightly lower rates at approximately 62\%. The palmvein model ResNet101  demonstrates comparable susceptibility with success rates around 61\%. Palmvein patterns are inherently harder to spoof due to their subtle, sub-surface nature, which makes them more resistant to EM leakage and reconstruction. This is different from the more prominent, surface-level features of palmprints that are easier to capture and exploit. These findings confirm that our EM-based reconstruction method poses a significant security threat across diverse models used in palm biometric systems.

An interesting observation is that among all palmprint models, PalmNet exhibits the highest vulnerability to \app{}. Unlike generic CNN-based models, PalmNet adopts a hybrid architecture that integrates Gabor filters with a PCA-based unsupervised scheme. This design choice makes PalmNet particularly susceptible to attacks from \app{}, as its strong capabilities at recovering principal textural features. This observation underscores a key insight:  models that depend heavily on low-level or principal-component-derived features may inadvertently expose themselves to greater risk when such features are recoverable through external leakage.
These findings highlight the need for model designs that are robust to side-channel reconstructions, potentially by avoiding over-reliance on easily reconstructible signal patterns and incorporating safeguards that account for fine-grained biometric information.

\begin{table}[h]
\centering
\caption{Effectiveness of \app{} on three COTS devices.}
\vspace{-10pt}
\label{tab:cots}
\resizebox{1.0\linewidth}{!}{
\begin{tabular}{l c c c c}
\toprule
\textbf{Device} & \textbf{SSIM} ↑ & \textbf{PSNR (dB)} ↑  & \textbf{FID}  ↓ & \textbf{Average SSR (\%)} ↑ \\\hline
C1 (Office Gate) & 0.64 & 27.8 & 11.3 & 52.5 \\\hline
C2 (Home Locker) & 0.61 & 26.4 & 11.7 & 59.1 \\\hline
C3 (Payment Kiosk) & 0.66 & 28.2 & 10.7 & 60.9 \\
\hline
\end{tabular}}
\vspace{-1em}
\end{table}

\noindent\textbf{Effectiveness of Attacking COTS Devices. }To further evaluate the practicality of \app{} in real-world settings, we extend our experiments to COTS palm recognition devices C1, C2, and C3, representing three typical deployment scenarios: \textit{Office Gate}, \textit{Home Locker}, and \textit{Payment Kiosk}.
% We select two representative COTS systems that are widely deployed in daily authentication scenarios. For security and confidentiality reasons, we do not disclose the exact models of these devices. 
We focus on assessing whether \app{} is effective on these commodity systems, in terms of its reconstruction quality and effectiveness in spoofing attacks. Table~\ref{tab:cots} reports the results of \app{} on the three COTS devices. Despite the differences in hardware design and shielding strategies, our results confirm that \app{} can successfully extract biometric information from COTS devices, with the reconstructed images demonstrating substantial spoofing capability against recognition models, highlighting the generality and severity of this threat.

\subsection{Impacts of Practical Factors}
\label{subsec:impact}

Unless otherwise specified, all impact experiments were conducted under a default configuration. The palm recognition software (PalmNet) and sensor models (V1 for palmprint, IR1 for palmvein) were used, with the sensor connected to the SBC (S1) under evaluation. A receiving antenna was placed at a fixed distance of 0.5~meter and paired with a 40~dB LNA to ensure sufficient signal strength.
 
\begin{table}[t]
\centering
\caption{Impact of different SBCs on \app{}.}
\vspace{-10pt}
\label{tab:SBCs}
\resizebox{1.0\linewidth}{!}{
\begin{tabular}{c c c c c}
\toprule
\textbf{Device} & \textbf{SSIM} ↑ & \textbf{PSNR (dB)} ↑  & \textbf{FID}  ↓ & \textbf{Average SSR (\%)} ↑ \\\hline
S1 & 0.72 & 29.41 & 8.73 & 62.7 \\\hline
S2 & 0.74 & 29.49 & 8.52 & 66.5 \\\hline
S3 & 0.72 & 29.24 & 9.12 & 60.1 \\
\bottomrule
\end{tabular}}
\end{table}

\noindent\textbf{Impact of Different SBCs. }To examine how different SBCs affect the performance of \app{}, we evaluated it on three single-board computers: Raspberry Pi 3B+ (S1), Raspberry Pi 5 (S2) and Jetson Nano (S3). Each device was configured with identical palmprint recognition software and connected to the same sensor model. The receiving antenna was placed at a fixed distance of 0.5 meters with 40 dB LNA, ensuring consistent experimental conditions across all tests. As reported in Table~\ref{tab:SBCs}, the performance of \app{} remains highly stable across different SBCs, confirming that the exploitable EM leakage originates from the sensor’s data transmission rather than the computing hardware. This demonstrates that the vulnerability is broadly applicable regardless of the deployment platform.

\noindent\textbf{Impact of Different LNAs. }To investigate the effect of low-noise amplifiers on \app{}, we conduct experiments using LNAs with different gain levels: no gain, 20dB, 30dB, and 40dB. The three gain levels correspond to different device models: ZK09-BM (20dB), Teyleten (30dB), and FST-RFAMP06 (40dB). 

\begin{figure}[t]
    \centering
    % 第一行
    \subcaptionbox{SSIM \label{fig:lnaSSIM}}{
        \includegraphics[width=0.47\linewidth]{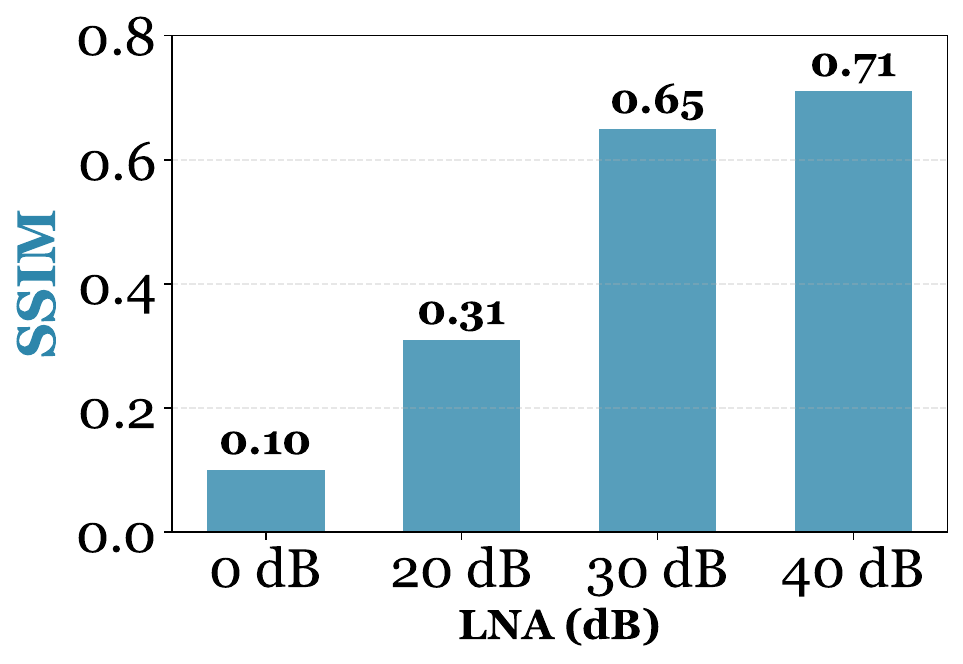}
    }\hfill
    \subcaptionbox{SSR \label{fig:lnaSSR}}{
        \includegraphics[width=0.47\linewidth]{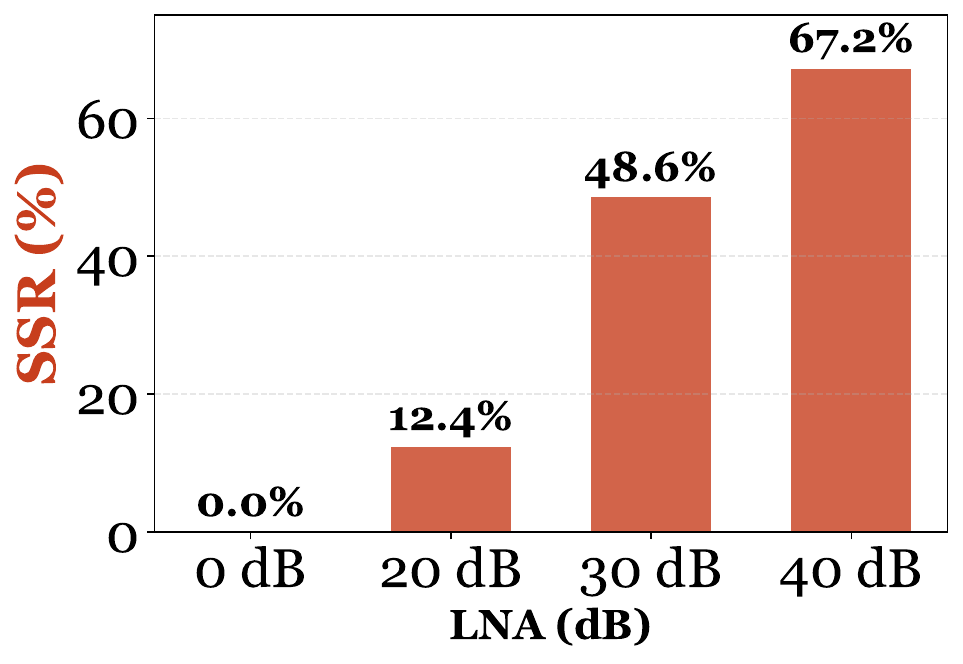}
    }
    \vspace{-1em}
    \caption{Impact of different LNAs on \app{}.}
    \label{fig:LNA}
    \Description{This figure shows ...}
    \vspace{-1em}
\end{figure}

\begin{figure}[t]
    \centering
    % 第一行
    \subcaptionbox{SSIM \label{fig:envSSIM}}{
        \includegraphics[width=0.47\linewidth]{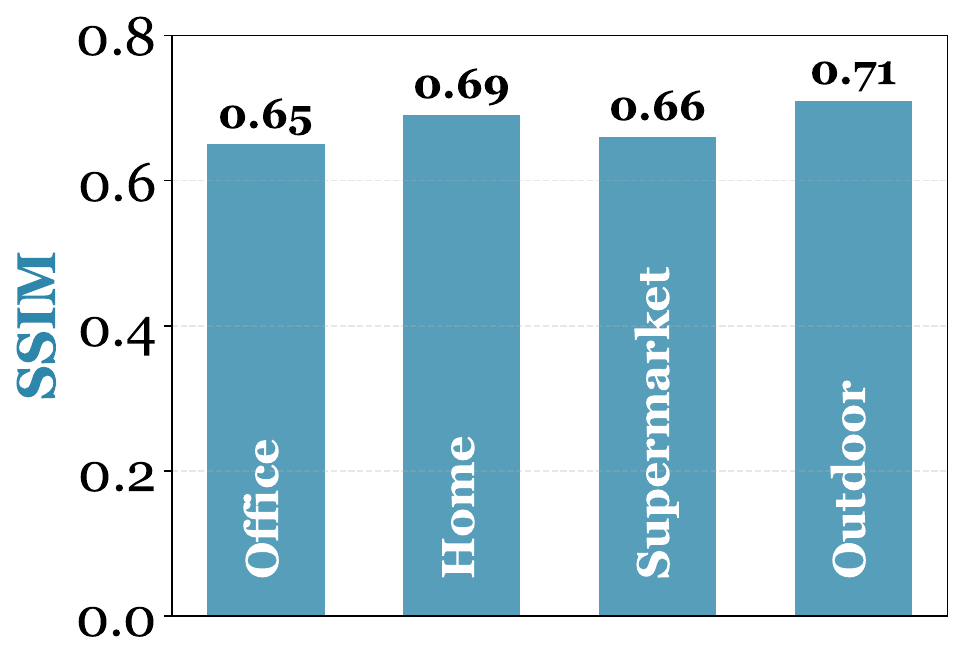}
    }\hfill
    \subcaptionbox{SSR \label{fig:envSSR}}{
        \includegraphics[width=0.47\linewidth]{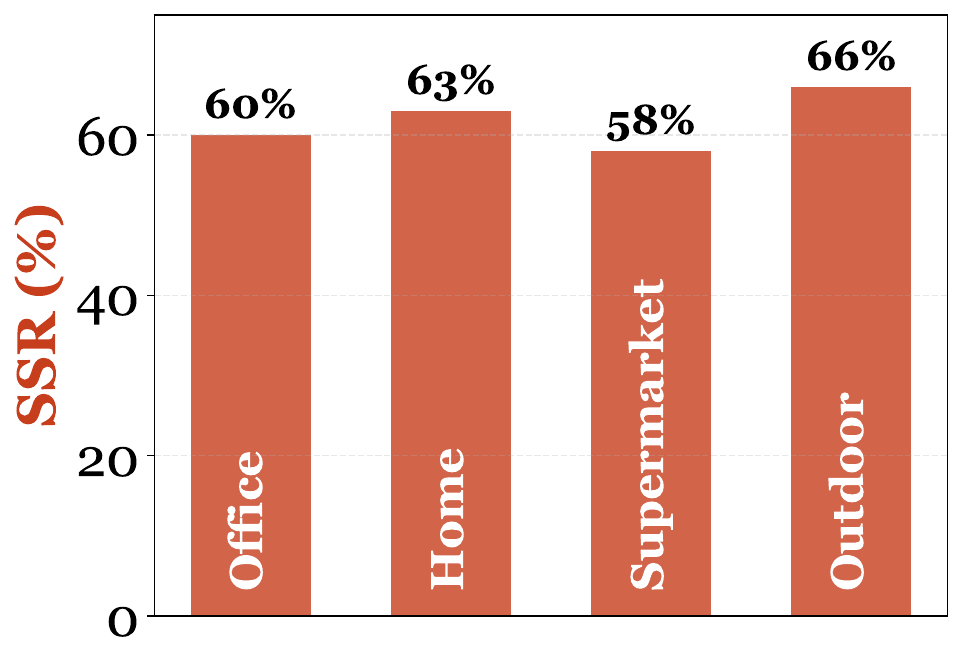}
    }
    \vspace{-1em}
    \caption{Impact of different environment noises.}
    \label{fig:env}
    \Description{This figure shows ...}
    \vspace{-1em}
\end{figure}

Figure~\ref{fig:LNA} presents the performance of \app{} across different LNA configurations. Without amplification (0dB), EM signals are too weak for meaningful palm restoration (SSIM < 0.1, SSR = 0\%). The 20dB amplifier shows minimal improvement (SSR = 12.4\%), remaining insufficient for practical attacks. However, substantial improvements emerge with 30dB amplification (SSIM = 0.65, SSR = 48.6\%), which further increase with the 40dB amplifier (SSIM = 0.71, SSR = 67.2\%). These results demonstrate a clear correlation between LNA gain and attack effectiveness, with a notable threshold effect between 20dB and 30dB, where the amplification becomes sufficient to capture fine-grained biometric features through EM emissions.

% \begin{figure}[t]
%     \centering
%     % 第一行
%     \subcaptionbox{SSIM \label{fig:ANGLESSIM}}{
%         \includegraphics[width=0.47\linewidth]{FIG/experiments/ANGLE_SSIM.pdf}
%     }\hfill
%     \subcaptionbox{SSR \label{fig:ANGlESSR}}{
%         \includegraphics[width=0.47\linewidth]{FIG/experiments/ANGLE_SSR.pdf}
%     }
%     \caption{Impact of antenna angle on \app{}. (a) SSIM and (b) SSR under varying angles $\theta$. Results indicate optimal reception zones and a complete signal loss at $\theta=90^\circ$.}
%     \label{fig:ANGLE}
% \end{figure}

\noindent\textbf{Impact of Different Environmental Noises. }To evaluate \app{}'s robustness against real-world noises, we tested the \app{} across four daily-life environments where palmprint recognition can be commonly deployed: office, home, supermarket, and outdoor settings. We maintained a fixed distance of 1.5 meters and collected 50 EM traces in each environment during peak activity hours to capture representative noise conditions. As shown in Figure~\ref{fig:env}, \app{} achieved consistent metrics across all environments, demonstrating the general effectiveness of \app{} in daily-life scenarios. This robustness to ambient interference validates EMPalm's practical threat potential in real-world deployments.

\begin{figure}[t]
    \centering
    \subcaptionbox{Directional antenna\label{fig:far_distance}}{%
        \includegraphics[width=0.48\linewidth]{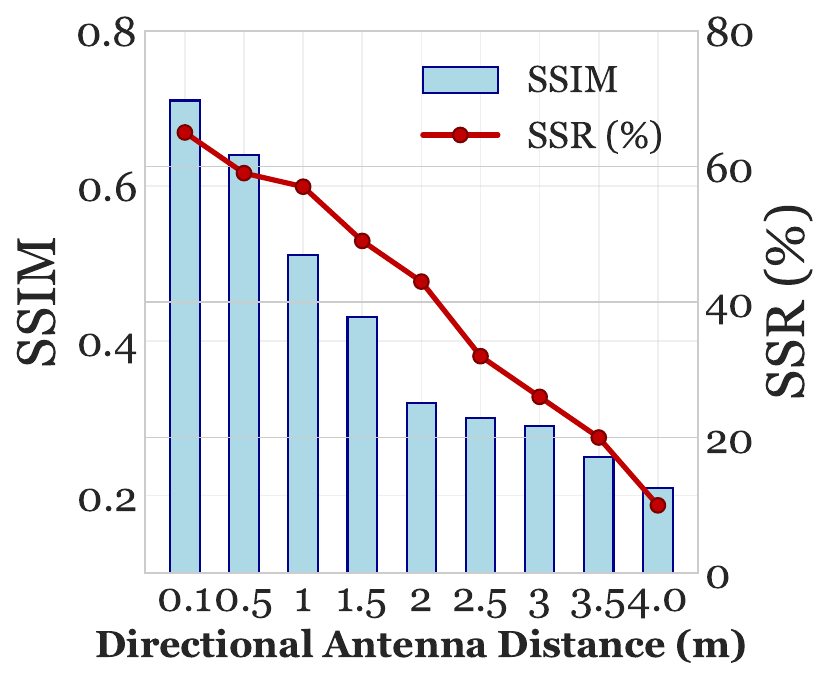}
    }\hfill
    \subcaptionbox{Nearby antenna\label{fig:near_distance}}{%
        \includegraphics[width=0.48\linewidth]{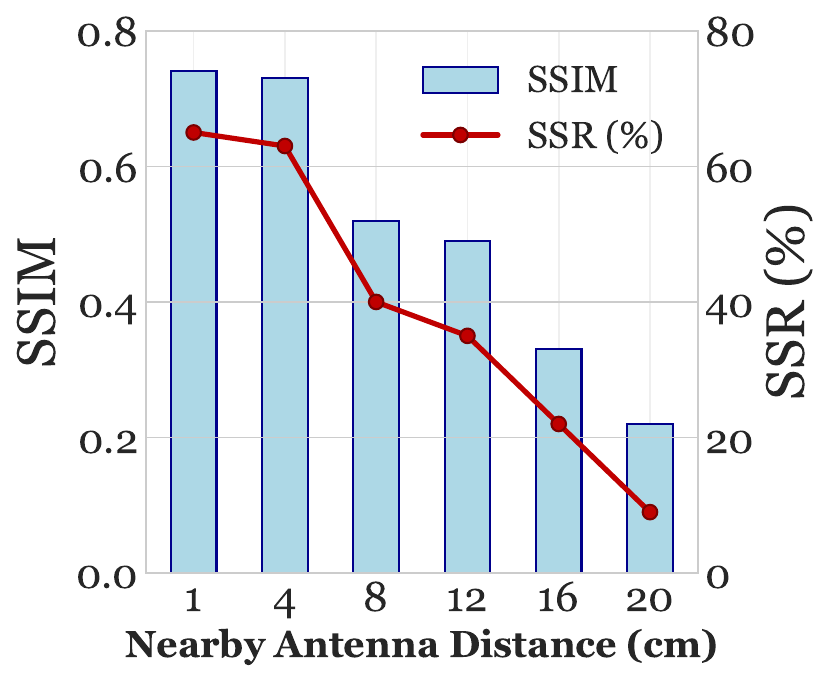}
    }
    \vspace{-8pt}
    \caption{Impact of antenna distance on \app{}.}
    \label{fig:impact_distance}
    \Description{This figure shows ...}
    \vspace{-1em}
\end{figure}
% \noindent\textbf{Impact of Different Distances. }We evaluated \app{}'s performance using directional antennas at distances from 0.1 meter to 4.0 meter with 0.5 meter intervals. As shown in Figure~\ref{fig:far_distance}, SSIM values decrease from 0.72 to 0.21 at 4.0 meters due to EM signal attenuation. The attack success rate (Figure~\ref{fig:far_distance}) also exhibits a significant decline, dropping from 65\% to 10\% at 4.0 meters.
% Notably, \app{} remains effective within a 2-meter range, achieving 43\% SSR and 0.32 SSIM at 2~meter—sufficient for practical attacks in common environments. Beyond 2 meters, performance gradually decreases due to multipath effects and low SNR, though directional antennas markedly extend the attack range, confirming EMPalm’s practical threat even from adjacent spaces.

\noindent\textbf{Impact of Different Distances. }
As shown in Figure~\ref{fig:impact_distance}, we evaluate \app{} under two antenna configurations. 
With a directional antenna (meter-level distances), performance gradually decreases from 0.72 SSIM and 65\% SSR at 0.1~m to 0.21 SSIM and 10\% SSR at 4.0~m due to EM attenuation. 
\app{} remains effective within 2~m, achieving 0.32 SSIM and 43\% SSR, indicating practical feasibility across typical room-scale environments.

With a nearby antenna (centimeter-level distances), performance remains consistently high within 4~cm, exceeding 0.7 SSIM and 60\% SSR, and then drops sharply beyond 8~cm as near-field coupling weakens. 
This contrast highlights strong short-range leakage and the extended reach enabled by directional antennas.

\begin{figure}[t]
    \centering
    \subcaptionbox{SSIM \label{fig:ANGLESSIM}}{
    \includegraphics[width=0.49\linewidth]{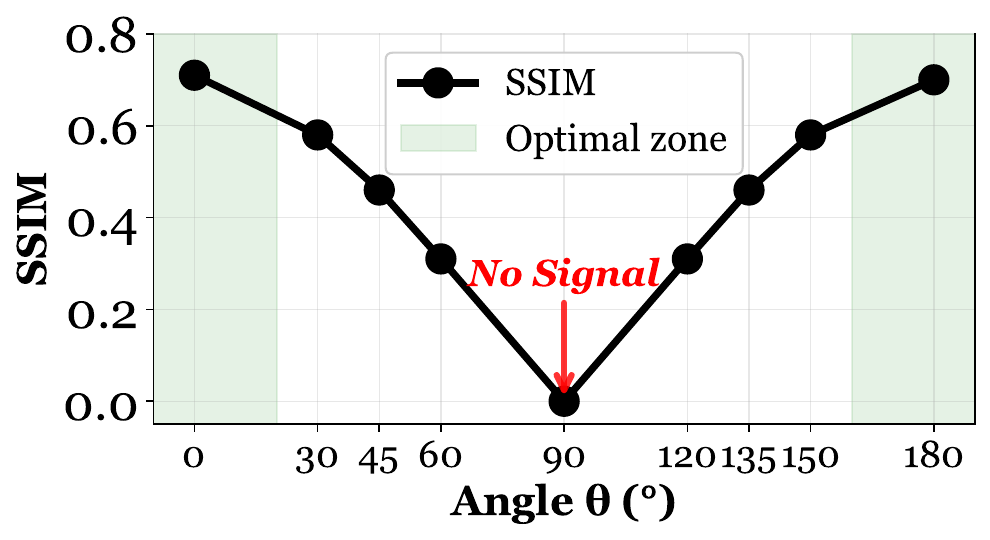}
    }\hfill
    \subcaptionbox{SSR\label{fig:ANGlESSR}}{\includegraphics[width=0.49\linewidth]{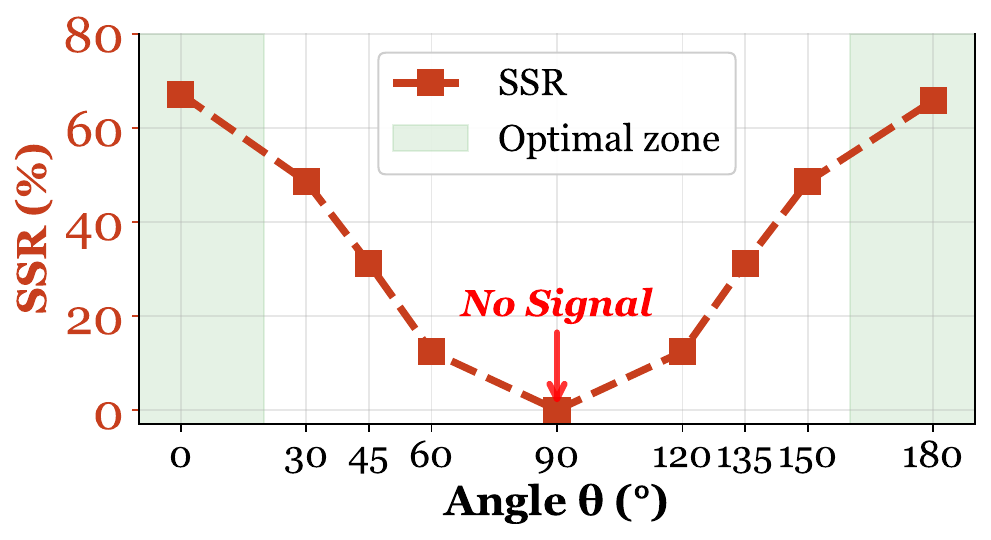}}
    \vspace{-10pt}
    \caption{Impact of antenna angle on \app{}. (a) SSIM and (b) SSR under varying angles $\theta$. Results indicate the presence of optimal reception zones at certain angles, while a complete signal loss is observed at $\theta=90^\circ$.}
    \label{fig:ANGLE}
    \Description{This figure shows ...}
    \vspace{-1em}
\end{figure}

\noindent\textbf{Impact of Different Probe Angles. }To evaluate the impact of probe orientation, we position the receiving probe 2~centimeters away from the target palm sensor in the near-field region and vary the probe angle $\theta$ from $0^{\circ}$ to $180^{\circ}$. Figure~\ref{fig:ANGLE} illustrates the relationship between probe angle and restoration quality measured by SSIM. The highest restoration quality is achieved when the receiving probe achieves optimal EM coupling with the sensor's internal signal paths at $0^{\circ}$ (SSIM = 0.71, SSR = 67.2\%) and $180^{\circ}$ (SSIM = 0.70, SSR = 65.8\%), where the probe orientation maximizes interception of the radiated EM fields. As the angle moves away from $0^{\circ}$/$180^{\circ}$, restoration quality drops, reaching moderate levels at $30^{\circ}$–$150^{\circ}$ and failing completely at $90^{\circ}$ (SSIM = 0, SSR = 0\%). The symmetric degradation pattern suggests dipole-like radiation characteristics, indicating a predictable angular dependency that adversaries can exploit to optimize interception and spoofing.

\noindent\textbf{Impact of Different Building Materials.}
We further evaluate \app{}’s robustness when EM signals pass through common building materials separating the palm recognition device from an eavesdropper. Each material is tested under typical deployment conditions without extra shielding. Table~\ref{tab:building_materials} presents the reconstruction and spoofing results across five representative materials, including wood, drywall, glass, concrete, and aluminum panel, along with a baseline case without obstruction. The unobstructed setup yields the best image quality (SSIM 0.71, PSNR 29.1 dB, FID 14.2, SSR 67.5\%). Non-conductive materials like wood, drywall, and glass cause moderate signal loss but still permit recognizable palm reconstruction. Dense or conductive materials such as concrete and aluminum panels strongly attenuate EM emissions, reducing image quality and spoofing success. Overall, building materials can weaken but not eliminate EM side-channel leakage.

\begin{table}[t]
\centering
\caption{Impact of common building materials on EM-Palm performance under practical deployment conditions.}
\vspace{-5pt}
\label{tab:building_materials}
\small
\begin{tabular}{lcccc}
\toprule
Material & SSIM $\uparrow$ & PSNR (dB) $\uparrow$ & FID $\downarrow$ & SSR (\%) $\uparrow$ \\
\midrule
No Building Material & 0.71 & 29.1 & 9.2 & 67.5 \\
\midrule
Wood     & 0.63 & 27.8 & 17.5 & 58.4 \\
Drywall     & 0.59 & 26.9 & 18.9 & 54.1 \\
Glass     & 0.56 & 25.7 & 20.3 & 50.2 \\
Concrete & 0.38 & 22.4 & 28.7 & 33.5 \\
Aluminum panel      & 0.35 & 21.8 & 30.1 & 29.7 \\
\bottomrule
\end{tabular}
\end{table}

\begin{figure}[t]
    \centering
    \subcaptionbox{SSIM \label{fig:shieldSSIM}}{
\includegraphics[width=0.48\linewidth]{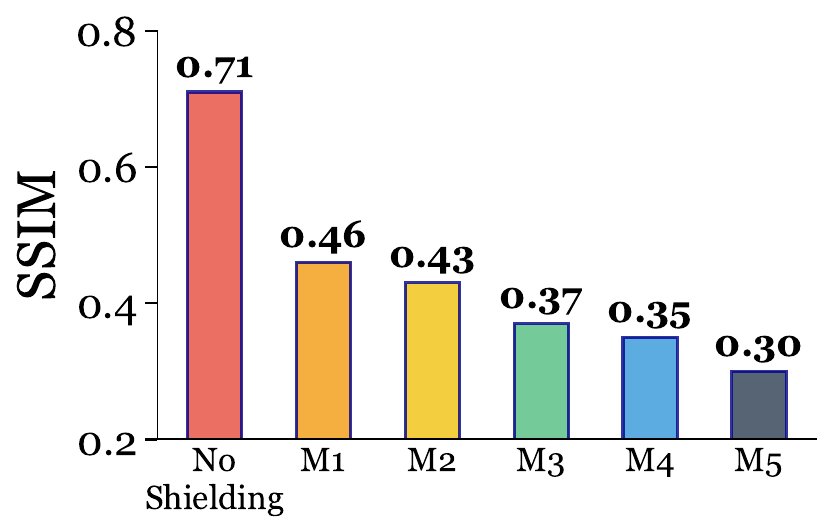}

    }\hfill
    \subcaptionbox{SSR\label{fig:shieldSSR}}{\includegraphics[width=0.48\linewidth]{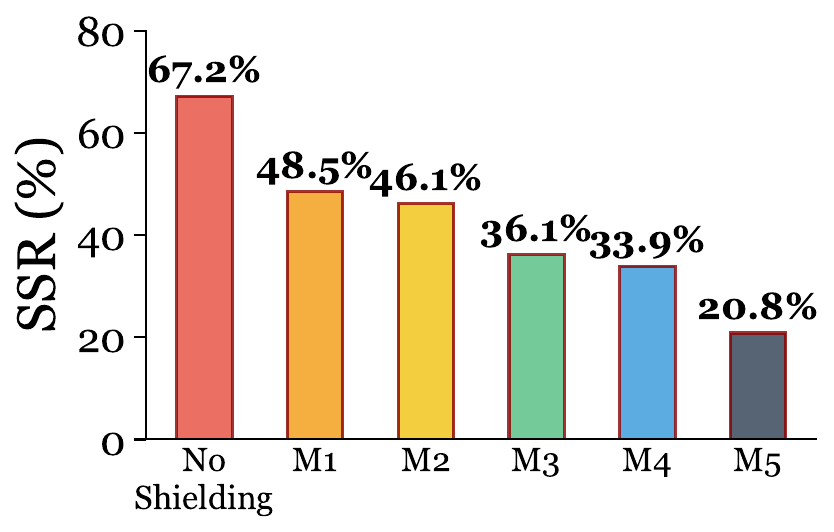}}
    \caption{Impact of EM shielding materials on \app{}. (a) SSIM and (b) SSR under varying EM shielding materials. }
    \vspace{-1em}
    \label{fig:EMshield}
    \Description{This figure shows ...}
    \vspace{-1em}
\end{figure}

\noindent\textbf{Impact of EM Shielding Materials. }Following EMIRIS~\cite{emiris} and EMeye~\cite{long2024eye}, we evaluated the impact of five shielding materials, including copper wire mesh (M5), aluminum foil (M4), metalized fabric (M3), conductive coating (M2), and conductive fabric (M1), on \app{}, with each material uniformly wrapped around the sensor’s data transmission cables. All other experimental settings remained identical to those described above. Figure~\ref{fig:EMshield} shows how different shielding materials affect \app{}, ordered by theoretical shielding capability. Conductive fabric (SSIM: 0.46, SSR: 48.5\%) and coating (SSIM: 0.43, SSR: 46.1\%) provide moderate protection. Metalized fabric (SSIM: 0.37, SSR: 36.1\%) and aluminum foil (SSIM: 0.35, SSR: 33.9\%) offer better suppression. Copper mesh delivers the strongest shielding (SSIM: 0.30, SSR: 20.8\%), significantly reducing reconstruction quality and spoofing success. These differences reflect variations in material conductivity, thickness, and structure. While EM shielding materials substantially degrade \app{}'s effectiveness, they cannot fully eliminate the side-channel vulnerability.

\begin{table}[t]
\centering
\caption{Cross-dataset validation results.}
\vspace{-10pt}
\label{tab:crossdataset}
\resizebox{\columnwidth}{!}{

\begin{tabular}{l l c c}
\toprule
\textbf{Configuration} & \textbf{Training Data} & \textbf{SSR (\%)} & \textbf{$\Delta$SSR} \\
\midrule
Original (50/50 split) & Split In CASIA+Tongji & 68.93 & - \\
\midrule
Separation Dataset  & Diff: Tongji only & \multirow{2}{*}{69.07} & \multirow{2}{*}{+ 0.14\%} \\
& Recog: CASIA only &  &  \\
\bottomrule
\end{tabular}
}
\end{table}

\noindent\textbf{Cross-Dataset Generalization. }To evaluate the generalization capability of \app{} across different data sources, we conduct a cross-dataset validation using completely disjoint datasets. Specifically, the diffusion model is trained only on Tongji, while the recognition model is trained only on CASIA, with no subject or data overlap. As shown in Table~\ref{tab:crossdataset}, the SSR remains comparable to the original 50/50 split setting, with only a marginal difference of $+0.14\%$, indicating that \app{} maintains stable performance when trained and evaluated on different datasets.

\noindent\textbf{Feasibility of Deferred Attack.}
Beyond real-time attacks, \app{} also supports deferred attack scenarios. An adversary can deploy a compact eavesdropping device to passively collect EM emissions over time and perform offline reconstruction once data is obtained. To validate this setting, we built a miniaturized collection system using an Adalm-Pluto~\cite{Pluto} and a microcontroller, with a footprint of only 6 $\times$ 10 $\times$ 3 centimeters, enabling discreet placement near palm recognition terminals. 
% We characterize its operational range and find the attack effective within 20 cm, beyond which signal strength degrades due to inverse-square attenuation. 
We collected In-phase and Quadrature (IQ) data from 10 authentication sessions over a 3-hour period using this compact setup. Notably, \app{} remains a single-shot attack, requiring EM emissions from only one authentication event (200–500 ms) to reconstruct. Offline analysis of stored IQ samples achieves reconstruction quality comparable to real-time attacks (Table~\ref{tab:extend}), demonstrating the practicality of long-term covert data collection combined with one-time attack execution.

\begin{table}[t]
\centering
\caption{Experimental results of offline palmprint reconstruction from IQ data collected using the compact setup (<20cm).}
\vspace{-10pt}
\label{tab:extend}
\resizebox{1.0\linewidth}{!}{
\begin{tabular}{c c c c c}
\toprule
\textbf{Users} & \textbf{SSIM} ↑ & \textbf{PSNR (dB)} ↑  & \textbf{FID}  ↓ & \textbf{Average SSR (\%)} ↑ \\\hline
U1 & 0.66 & 26.18 & 9.94 & 53.6 \\\hline
U2 & 0.70 & 26.44 & 9.62 & 57.1 \\
\hline
\end{tabular}}
\vspace{-1em}
\end{table}

\subsection{Ablation Study}
To quantify the individual contribution of each module in \app{}, we conduct comprehensive ablation experiments.
Each variant removes one key component while keeping the rest
of the pipeline unchanged.
%Specifically, 
We construct four leave-one-out variants:  
(i) \textit{without Dual-Modal Disentanglement}, which reconstructs mixed palmprint/palmvein signals directly; 
(ii) \textit{without Multi-Band Combination}, which uses only the single best-SNR band; and 
(iii) \textit{without DiffPIR Restoration}, which omits the restoration module.

Table~\ref{tab:ablation} summarizes the ablation results. 
Removing any individual component substantially degrades both reconstruction fidelity and SSR, confirming that all modules are essential to \app{}. 
In particular, disabling Dual-Modal Disentanglement leads to the most severe structural collapse (SSIM drops from 0.74 to 0.25) and reduces SSR to below the effective detection threshold (<10\%),
indicating a practical attack failure, highlighting its critical role in preserving identity-separable features. Removing Multi-Band Combination causes a moderate performance drop (SSIM decreases to 0.54 and SSR to 52.8\%), suggesting that multi-band fusion enhances stability but is not the primary performance driver.
Moreover, removing DiffPIR Restoration dramatically increases perceptual distortion (FID rises from 8.6 to 19.4) and reduces SSR to 23.8\%, demonstrating that restoration is indispensable for recovering texture details.

\noindent\textbf{Per-stage Error Analysis. }Table~\ref{tab:stage} reports the cumulative error across the \app{} pipeline. 
Reconstruction fidelity improves progressively from raw reconstruction to multi-band fusion and finally to restoration. 
Notably, while structural metrics increase steadily, SSR exhibits a significant improvement after the restoration stage, indicating that generative refinement is critical for recovering identity-discriminative details.

\begin{table}[t]
\centering
\caption{Ablation study of \app{}.}
\vspace{-10pt}
\label{tab:ablation}
\resizebox{1.0\linewidth}{!}{
\begin{tabular}{l c c c c}
\toprule
\textbf{Variant} & \textbf{SSIM} $\uparrow$ & \textbf{PSNR (dB)} $\uparrow$ & \textbf{FID} $\downarrow$ & \textbf{SSR (\%)} $\uparrow$ \\
\hline
Full EMPalm & 0.74 & 29.5 & 8.6 & 66.5 \\
\hline
w/o Dual-Modal Disentanglement 
& 0.25 {\color{red}(-0.49↓)} 
& 14.7 {\color{red}(-14.8↓)} 
& 15.7 {\color{red}(+7.1↑)} 
& -- \\

w/o Multi-Band Combination 
& 0.54 {\color{red}(-0.20↓)} 
& 25.9 {\color{red}(-3.6↓)} 
& 10.6 {\color{red}(+2.0↑)} 
& 52.8 {\color{red}(-13.7↓)} \\

w/o DiffPIR Restoration 
& 0.52 {\color{red}(-0.22↓)} 
& 20.8 {\color{red}(-8.7↓)} 
& 19.4 {\color{red}(+10.8↑)} 
& 23.8 {\color{red}(-42.7↓)} \\
\hline
\end{tabular}}
\vspace{-1em}
\end{table}

\begin{table}[t]
\centering
\caption{Per-stage error analysis of \app{}.}
\vspace{-10pt}
\label{tab:stage}
\resizebox{1.0\linewidth}{!}{
\begin{tabular}{l c c c c}
\toprule
\textbf{Stage} & \textbf{SSIM} $\uparrow$ & \textbf{PSNR (dB)} $\uparrow$ & \textbf{FID} $\downarrow$ & \textbf{SSR (\%)} $\uparrow$ \\
\hline
Raw Reconstruction & 0.46 & 16.7 & 28.8 & -- \\
\hline
+ Multi-Band Combination 
& 0.52 {\color{darkgreen}(+0.06↑)} 
& 20.8 {\color{darkgreen}(+4.1↑)} 
& 19.4 {\color{darkgreen}(-9.4↓)} 
& 23.8 {\color{darkgreen}(+23.8↑)} \\
\hline
+ DiffPIR Restoration (Full) 
& 0.74 {\color{darkgreen}(+0.28↑)} 
& 29.5 {\color{darkgreen}(+12.8↑)} 
& 8.6 {\color{darkgreen}(-20.2↓)} 
& 66.5 {\color{darkgreen}(+66.5↑)} \\
\hline
\end{tabular}}
\raggedright\footnotesize{\textit{Note: SSR values below 10\% are denoted as “–”, as they indicate near-random attack performance and are considered negligible.}}
\vspace{-1em}
\end{table}

\section{Discussion}

% \noindent\textbf{Real-world Attack Implications. }Beyond digital spoofing, the reconstructed biometric data from \app{} enables physical attacks. Researchers at the Chaos Communication Congress demonstrated that wax hand models containing vein patterns can fool commercial authentication systems~\cite{ccc2023}. By combining \app{} with such fabrication techniques, adversaries can execute the complete attack chain—from covert EM collection to physical access—without ever contacting the victim. This fundamentally undermines the security of palmprint and palmvein authentication systems deployed in critical infrastructure.

\noindent\textbf{Countermeasures.}
Based on the vulnerabilities identified in Section~\ref{subsec: Reconstruction}, several defenses can mitigate the risks posed by \app{}. 
First, EM shielding applied to sensor transmission cables can suppress informative emissions, and appropriate material choices can significantly reduce reconstruction quality and spoofing success.
Second, redesigning the transmission protocol—such as increasing transmission complexity or decoupling packets from pixel-level information—can break the direct mapping between EM signals and biometric data.
Third, system-level defenses, including anomaly detection and multi-factor authentication, can help prevent spoofing using reconstructed or fabricated artifacts.
Together, these measures form a multi-layered defense strategy spanning hardware shielding, protocol hardening, and system-level security enhancements against EM side-channel attacks.

% \noindent\textbf{Responsible Disclosure.}  
% We have notified the corresponding vendors of the discovered security vulnerabilities and provided them with potential countermeasures. This process was conducted in accordance with the principles of responsible disclosure.

\noindent\textbf{Limitations and Future Work.} Palm recognition sensors are embedded in complex electronic systems, where EM emissions from surrounding components, channel effects, and hardware imperfections inevitably introduce interference and reconstruction errors. Limited sampling rate and bandwidth further attenuate high-frequency details, and polarity inversion may cause grayscale distortion, constraining fine-grained reconstruction and effective attack range. While our approach reliably recovers the key palmprint and palmvein structures required for physical spoofing and deceiving most image-based commercial systems, its effectiveness against high-end devices with liveness detection or enhanced protection mechanisms remains an open question. Future work will focus on improving EM signal processing to enhance SNR and suppress interference, exploring advanced antenna and denoising techniques to better characterize and potentially extend effective attack distances (e.g., via directional beamforming or learning-based signal enhancement), and fabricating realistic 3D prosthetic hands~\cite{ccc2023} to validate end-to-end physical spoofing.

\section{Related Work}
\noindent\textbf{Diverse EM Side-channel Attack Surfaces.} Prior works have demonstrated the exploitation of EM side channels across diverse systems, including keystroke and browsing reconstruction from GPUs~\cite{GPU}, fingerprint recovery from in-display fingerprint sensors~\cite{ni2023finger}, high-fidelity iris reconstruction from NIR sensors~\cite{emiris}, and video stream extraction from embedded cameras~\cite{long2024eye}. Research has further shown that smartphone magnetometers can analyze EM footprints to infer running applications~\cite{zhu2016,pan2021magthief}, while wireless charging inadvertently leaks sensitive information through EM emissions~\cite{ni2023exploiting,li2023magfingerprint,wireless}. Additional studies have revealed EM vulnerabilities in cryptographic implementations~\cite{cheng2019}, smartphone activity inference~\cite{fu2024magspy}, USB device fingerprinting~\cite{usb}, hidden camera detection~\cite{liu2023camradar}, hidden microphone detection~\cite{zhou2023dehirec} and IoT activity profiling~\cite{iot,zhao2025membership,liu2024emtrig,xu2025magwatch}. EM analysis has also been extended to system- and device-level threats, such as detecting GPU cryptojacking via magnetic leakage (MagTracer)\cite{hanjun1} and identifying laptop microphone recording states through EM emanations (TickTock)\cite{hanjun2}.

Overall, \app{} advances prior work by (i) providing the first empirical exploitation of EM leakage in palm recognition with simultaneous disentanglement of interleaved palmprint and palmvein signals, and (ii) enabling EM-based palm image restoration using an unsupervised diffusion framework trained solely on public palm datasets, without requiring paired EM data or device-specific calibration. Together, \app{} offers a unified framework linking EM leakage analysis to practical biometric exploitation.

% Beyond passive eavesdropping, recent research has demonstrated that intentional electromagnetic interference (IEMI) can serve as a potent non-contact attack vector against sensing and imaging systems. GlitchHiker\cite{jiang2023glitchhiker} shows that IEMI can disrupt camera image-signal transmission, inducing controlled glitches that manipulate captured frames. GhostType\cite{jiang2024ghosttype} demonstrates that IEMI can inject phantom keystrokes into keyboard circuits, enabling large-scale false input or denial-of-service attacks. GhostShot~\cite{ren2025ghostshot} further reveals that IEMI can inject arbitrary grayscale or colored patterns into CCD cameras, thereby falsifying computer vision outputs under normal lighting conditions.

\section{Conclusion}
In this paper, we propose \app{}, the first EM side-channel attack recovering palm biometrics from recognition systems. \app{} handles both single- and dual-modality systems by reverse-engineering transmission protocols and employing three techniques: frame boundary identification with modality disentanglement, multi-band image combination for bit recovery, and DiffPIR-based texture restoration. Our experiments show that \app{} reconstructs high-fidelity palm images from EM signals, exhibiting strong structural similarity, high signal quality, and low perceptual discrepancy, as well as enabling successful spoofing across diverse recognition models. These findings reveal critical vulnerabilities in existing palm recognition systems, stressing the importance of using improved multi-factor defenses for better security.

\begin{acks}
This work is supported by NSF DGE-2409851. Xiaoyan Sun and Jun Dai are also supported by NSF OAC-2528534.
\end{acks}


% \begin{acks}
% \end{acks}

\bibliographystyle{ACM-Reference-Format}
\bibliography{Ref}


\end{document}